\documentclass[aps,notitlepage,twocolumn,prl,nolongbibliography,superscriptaddress]{revtex4-2}
\usepackage{graphicx}
\usepackage{float}
\usepackage{amsmath}
\usepackage{amssymb}
\usepackage{hyperref}
\usepackage{bm}
\usepackage{color}
\usepackage{siunitx}

\newcommand{\thermal}[1]{\langle #1 \rangle}
\newcommand{\Tr}{\operatorname{Tr}}
\newcommand{\nint}{\operatorname{nint}}

\begin{document}

%\linenumbers
\title{Quantum Brownian Motion: proving that the Schmid transition belongs to the Berezinskii–Kosterlitz–Thouless universality class}

\author{F. G. Capone}
\affiliation{Dipartimento di Fisica ``E. Pancini'', Universit\`a di Napoli Federico II,
Complesso Universitario di Monte Sant'Angelo, via Cintia, 80126 Napoli, Italy}
\affiliation{INFN, Sezione di Napoli, Napoli, Italy}

\author{A. de Candia}
\affiliation{SPIN-CNR and Dip. di Fisica E. Pancini - Università di Napoli Federico II - I-80126 Napoli, Italy}
\affiliation{INFN, Sezione di Napoli, Napoli, Italy}

\author{V. Cataudella}
\affiliation{SPIN-CNR and Dip. di Fisica E. Pancini - Università di Napoli Federico II - I-80126 Napoli, Italy}
\affiliation{INFN, Sezione di Napoli, Napoli, Italy}

\author{R. Fazio}
\affiliation{Dipartimento di Fisica ``E. Pancini'', Universit\`a di Napoli Federico II,
Complesso Universitario di Monte Sant'Angelo, via Cintia, 80126 Napoli, Italy}
\affiliation{The Abdus Salam International Center for Theoretical Physics (ICTP), Strada Costiera 11, 34151 Trieste, Italy}

\author{N. Nagaosa}
\affiliation{RIKEN Center for Emergent Matter Science (CEMS), Wako, Saitama 351-0198, Japan}
\affiliation{Fundamental Quantum Science Program (FQSP), TRIP Headquarters, RIKEN, Wako 351-0198, Japan}

\author{C. A. Perroni}
\affiliation{SPIN-CNR and Dip. di Fisica E. Pancini - Università di Napoli Federico II - I-80126 Napoli, Italy}
\affiliation{INFN, Sezione di Napoli, Napoli, Italy}

\author{G. De Filippis}
\affiliation{SPIN-CNR and Dip. di Fisica E. Pancini - Università di Napoli Federico II - I-80126 Napoli, Italy}
\affiliation{INFN, Sezione di Napoli, Napoli, Italy}

\begin{abstract}
We investigate the equilibrium properties of a quantum Brownian particle moving in a periodic potential, specifically addressing the nature of the dissipation-driven Schmid transition in the Ohmic regime. By employing World-Line Monte Carlo in the path-integral formalism and introducing a specific binary order parameter, we demonstrate that the transition belongs to the Berezinskii-Kosterlitz-Thouless universality class. This classification is substantiated through finite-size scaling analysis that reveals the characteristic logarithmic decay of the correlation functions associated with the order parameter at the critical point. Quantum phase transition turns out to be extremely fragile: it disappears in both over- and sub-Ohmic dissipation regimes. Crucially, we find that the presence of the periodic potential does not alter the localization properties in the sub-Ohmic and super-Ohmic regimes, where the system exhibits the same qualitative behavior as the free quantum Brownian particle. These findings highlight that the emergence of critical behavior is strictly governed by the low-frequency form of the environmental spectral function, which determines the long-range temporal decay of the dissipative kernel.
\end{abstract}

\maketitle

\textit{Introduction.---}
Quantum technologies are gaining increasing attention for their potential impact on communication, simulation, computation, sensing, and metrology~\cite{Dowling_rev, Acin_rev}. Their practical development is challenging, as it requires controlling quantum systems while shielding them from environmental disturbances. Understanding open quantum systems is therefore essential for designing devices that operate reliably in realistic conditions.~\cite{Zurek_rev,breuer_petruccione_book}.

The quantum Brownian motion (QBM) model, the quantum analog of the classical Brownian motion, provides the most successful framework for studying a particle coupled to an environmental bath~\cite{breuer_petruccione_book, Caldeira_Leggett_1, Grabert}. In particular, QBM in a periodic potential acquires huge relevance for Josephson junction description and is the natural arena to study tunneling, dissipation, and the competition between quantum coherence and environmental coupling~\cite{Caldeira_Leggett_2, Chakravarty}. 
Lowering the temperature enhances coherent tunneling between adjacent minima of the periodic potential, whereas increasing the system--bath coupling promotes localization and decoherence, driving the particle to behave classically and become trapped in a single well. This competition suggests the possibility of a zero--temperature dissipation-driven localization-delocalization quantum phase transition (QPT)~\cite{Sachdev_book}. 

Ohmic dissipation has been extensively investigated in both the past~\cite{Schmid, Bulgadaev, Fisher_Zwerger, Weiss_5th_edition} and more recently~\cite{Troyer_MC, Troyer_PRL, Comment_absence_Hakonen_Sonin, Glazman_1, Glazman_2, Glazman_3, Glazman_4, Glazman_5, Glazman_6}. These works predicted that the system undergoes the so-called Schmid transition \cite{Schmid, Bulgadaev}, a dissipative QPT whose critical point is independent of the ratio between the amplitude $E_J$ of the cosine potential and the kinetic energy scale $\hbar^2 k^2/(2C)$, with $k$ the inverse period of the potential and $C$ the mass of the particle. In the case of a resistively shunted Josephson junction (RSJJ), the localization-delocalization transition of the QBM has also been predicted to give rise to an insulating-superconducting transition in the dc response, where the critical point is determined by the quantum resistance $R_Q = h/(4e^2)$; for shunt resistances below $R_Q$, the system should be described by the superconducting state. This paradigm has recently attracted renewed attention. Experimentally, it has been argued \cite{Murani_absence, Murani_reply} that there is no signature of critical behavior on the transport properties of the RSJJ, concluding that the system is always superconducting independently of the value of the shunt resistance. Concurrently, the very existence of the transition has been challenged on theoretical grounds \cite{Murani_absence, Murani_reply, Altimiras:2023trq, notaa, Masuki_Absence, Masuki_reply, Comment_absence_Sepulcre}. 

The main aim of this letter is to answer the following questions: 1) does the QPT predicted by Schmid~\cite{Schmid} and Bulgadaev~\cite{Bulgadaev} occur? 2) If yes, what is its nature?
By employing a numerically exact World-Line Monte Carlo (WLMC) in imaginary time, a variant of the technique introduced in Refs.~\cite{Troyer_MC, Troyer_PRL, SuppMatRef}, we analyze the presence and investigate the nature of the localization-delocalization transition for a QBM with a periodic potential and identify which kinds of coupling with the environment and periodic potential combine to produce genuine critical behavior. 

\textit{The model.---}
We consider the Hamiltonian
\begin{equation}\label{eqn:hamiltonian}
            H = \frac{Q^2}{2C} - E_J\cos\biggl(\frac{\phi}{\phi_0}2 \pi\biggr)  + \sum_{j=1}^N \biggl[ \frac{q_j^2}{2C_j} + \frac{1}{2L_j}\bigl(\phi_j -\phi\bigr)^2\biggr],
\end{equation}
where the canonical commutation relations $[\phi,Q] = i\hbar$ and $[\phi_j,q_k] = i\hbar\delta_{jk}$ hold. A general interpretation of Hamiltonian~\eqref{eqn:hamiltonian} is that of a particle with coordinate $\phi$, momentum $Q$, and effective mass $C$, moving in a cosine potential of amplitude $E_J$ and spatial periodicity $\phi_0$ (the system), while linearly coupled to a large set of $N$ harmonic oscillators (the environment).

We introduce the partition function of the system, $Z = \Tr e^{-\beta H}$, at temperature $T = (k_B\beta)^{-1}$.
In the imaginary-time path integral formalism, integrating out the environmental degrees of freedom yields
\begin{equation}\label{eqn:partition_path}
            Z = \int\limits_{\varphi(0)=\varphi(\beta\hbar)} {\cal D}\bigl[\varphi(\tau)\bigr] \,
            e^{-{\cal S}[\varphi(\tau)]/\hbar},
\end{equation}
where the effective Euclidean action reads
\begin{equation}\begin{split}\label{eqn:action}
            {\cal S}[\varphi(\tau)] = &\frac{\hbar^2}{4E_C}\int\limits_0^{\beta \hbar} d\tau \, \biggl(\frac{d\varphi}{d\tau}\biggr)^2 
            - E_J \int\limits_0^{\beta \hbar} d\tau \, \cos \varphi(\tau) \\
            &+ \int\limits_0^{\beta \hbar} d\tau \int\limits_0^{\beta \hbar} d\tau' \,
            K(\tau-\tau') \bigl(\varphi(\tau)-\varphi(\tau')\bigr)^2.
\end{split}
\end{equation}
Here, $E_C = h^2/(2C\phi_0^2)$ and $\varphi = 2\pi\phi/\phi_0$ is the dimensionless coordinate. The influence of the environment is fully encapsulated in the kernel function
\begin{equation}\label{eqn:kernel}
            K(\tau) = \frac{\hbar}{8\pi^2}\int\limits_0^{+\infty} d\omega \, J(\omega)\frac{\cosh\bigl[\omega(\beta\hbar/2 -|\tau|)\bigr]}{\sinh(\beta\hbar\omega/2)},
\end{equation}
which is determined by the spectral density
\begin{equation}\label{eqn:spectral}
            J(\omega) = \frac{\pi}{2}\frac{\phi_0^2}{h}\sum_{j=1}^N\frac{\omega_j}{L_j} \delta(\omega-\omega_j).
\end{equation}
Hence, it is clear that the physics depends only on the behavior of the function $J(\omega)$. In the following, we will consider a phenomenological power-law form
\begin{equation}\label{eqn:spectral_form}
            J(\omega) = \alpha \, \omega^s \, \omega_c^{1-s}\,\Theta(\omega_c-\omega).
\end{equation}
Dissipation turns out to be \textit{Ohmic} if $s=1$, \textit{sub-Ohmic} if $0<s<1$, and \textit{super-Ohmic} if $s>1$. The dimensionless parameter $\alpha$ controls the coupling with the environment, while $\omega_c$ is a cutoff frequency. The behavior of the retarded interaction, described by the last contribution of Eq.\eqref{eqn:action}, is determined by the expression of the spectral function at low frequencies. Indeed, it is straightforward to prove that $K(\tau) \sim \tau^{-1-s}$ for large $\tau$, provided $\tau < \beta \hbar$. In particular, we emphasize that in the Ohmic regime, i.e. $s=1$, $K(\tau)$ exhibits a quadratic power-law decay. 

We will prove that only this asymptotic temporal decay ensures the presence of a quantum phase transition by varying $\alpha$. To this aim, we will also take into account a modified spectral density:
\begin{equation}\label{eqn:spectral_form2}
J(\omega) = 
\begin{cases}
\alpha\omega & {\tilde \omega} <\omega<\omega_c, \\
\alpha{\tilde\omega}^{1-s} \, \omega^s & 0<\omega<{\tilde\omega},
\end{cases}        
\end{equation}
being $\tilde{\omega}$ a lower cut-off frequency. It introduces a deviation from ${\tau^{-2}}$ decay of $K(\tau)$ only at large times, of the order of $1/\tilde{\omega}$.

We highlight that if $s=1$ and $\phi_0 = h/(2e)$, Eqs.~\eqref{eqn:hamiltonian} and~\eqref{eqn:action} describe an RSJJ characterized by a charging energy $E_C$ and a Josephson energy $E_J$, with Ohmic resistance $R_S = R_Q/\alpha$. In this case, $\varphi$ denotes the phase difference operator between the two superconductors and $Q$ represents the charge on the capacitor plate~\cite{Vool-Devoret, Schon_Zaikin_review}. It is worth noting that even within this interpretation the phase difference operator must be understood in its \textit{extended picture}~\cite{Likharev,Averin,Comment_absence_Hakonen_Sonin}, with eigenvalues spanning the entire real axis. 

Throughout this work, we set $\hbar\omega_c = 50E_C$ and use $E_C=1$ as our reference energy scale. For continuous variable models, WLMC require a discretized form of the action~\eqref{eqn:action}. Specifically, we set $\Delta \tau = \frac{2\hbar}{3E_C}$. It should be noted that this choice does not affect the long-time properties of the dissipative kernel $K(\tau)$, which is the only source of a critical behavior at $T=0$ and thus does not modify the presence nor absence of the transition, nor its universality class, if present. In the latter case, only a slight renormalization of the critical point is expected.

\textit{Vanishing cosine potential: analytic results.---}
Let us pause concisely on the limiting case $E_J/E_C = 0$, where the action \eqref{eqn:action} is quadratic and thus the partition function can be exactly calculated. This case is of particular interest. Indeed, while for Ohmic dissipation ($s=1$) the Schmid transition has been predicted to occur at $\alpha_c=1$ for any $E_J/E_C > 0$ (in the $\omega_c \to \infty$ and $\Delta \tau \to 0$ limits), we show that at $E_J/E_C = 0$ no QPT takes place for any $s > 0$ and $\alpha > 0$. This result signals a non-analyticity in the phase diagram in the Ohmic regime.

The key quantity that proves the presence of a dissipation--driven QPT at a critical value of $\alpha$, i.e. $\alpha_c$, is $\sigma^2 \equiv \langle\bigl(\varphi -\bar\varphi\bigr)^2\rangle$, representing the variance of the dimensionless coordinate, where $\bar \varphi = \frac{1}{\beta\hbar}\int_0^{\beta \hbar} d\tau \, \varphi(\tau)$. It signals the change from a delocalized phase, i.e. $\sigma^2 \to \infty$ as $\beta \to \infty$ for $\alpha<\alpha_c$, to a localized one for $\alpha>\alpha_c$, i.e. $\sigma^2$ approaching a constant as $\beta\to \infty$.
\begin{figure}
\centering
\includegraphics[width=\columnwidth]{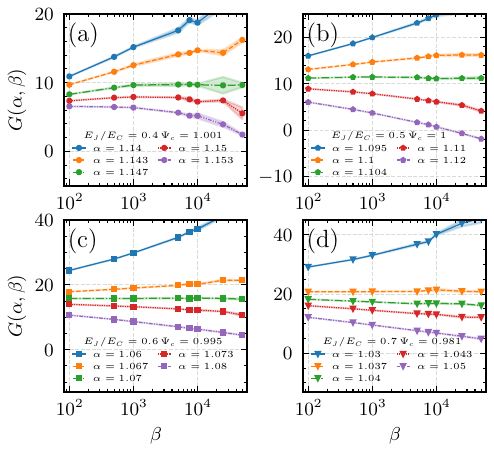}
\caption{Scaling function $G(\alpha, \beta)$ as a function of $\beta$ (in units of $1/E_C$) for $E_J/E_C=0.4$ (a), $E_J/E_C=0.5$ (b), $E_J/E_C=0.6$ (c), $E_J/E_C=0.7$ (d), for different values of the coupling $\alpha$ around the expected critical point for the Schmid transition. Estimated critical couplings are $\alpha_c \approx 1.147$, $\alpha_c \approx 1.104$, $\alpha_c \approx 1.07$, $\alpha_c \approx 1.037$ respectively. The legends indicate the value of the jump $\Psi_c$ used to generate the curves.
}
\label{fig:minn}
\end{figure}
In the Matsubara formalism, the variance is expressed as $\sigma^2 = \frac{2}{(\beta \hbar)^2}\sum_{m =1}^{\infty} \langle| \tilde{\varphi}_{i\omega_m}|^2\rangle$. Therefore, using the well known results for Gaussian integrals, it is exactly calculated as \cite{EndMatterRef}
\begin{equation}\begin{split}{\label{eqn:variance_Ej0}}
     \sigma^2 = \frac{1}{\beta\hbar} \sum_{m=1}^{\infty}\frac{\hbar}{\frac{\hbar}{4E_C}\omega_m^2+2\bar{K}_{i\omega_m}}.
\end{split}
\end{equation}
It should be emphasized that ${\bar K}_{i\omega_m} \equiv {\tilde K}_{i\omega_l=0}-{\tilde K}_{i\omega_m} = \int_0^{\beta \hbar} d\tau \, K(\tau)\bigl(1-\cos(\omega_m\tau)\bigr)$.
To obtain the asymptotic behavior of $\sigma^2$ for $\beta \to \infty$, we approximate the sum on the right-hand side with an integral, as
\begin{equation}\begin{split}{\label{eqn:asymptotic_Ej0}}
     \sigma^2 \approx \int\limits_{2\pi/(\beta\hbar)}^{\infty}\frac{d\omega}{2\pi}\,\frac{\hbar}{\frac{\hbar}{4E_C}\omega^2+2\bar{K}( i\omega)},
\end{split}
\end{equation}
\begin{figure}
\centering
\includegraphics[width=\columnwidth]{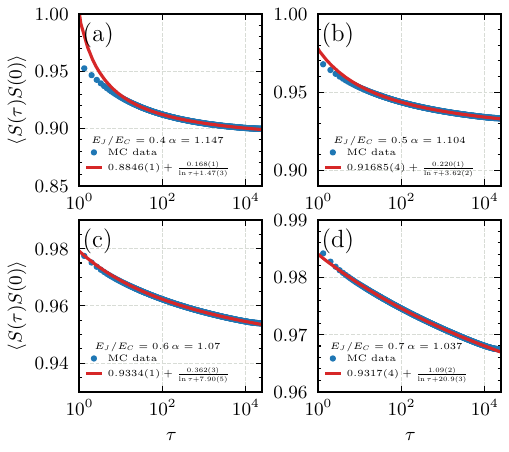}
\caption{Correlation function $\langle S(\tau)S(0)\rangle$ (blue dots) and corresponding logarithmic fit (red line) as functions of the imaginary time (in units of $\hbar/E_C$) for $E_J/E_C=0.4$ (a), $E_J/E_C=0.5$ (b), $E_J/E_C=0.6$ (c), $E_J/E_C=0.7$ (d), at $\beta E_C=50000$ for critical couplings $\alpha_c$ estimated via $G(\alpha, \beta)$ scaling argument. It should be noted that deviations from logarithmic decay only appear at short time and at $\tau \approx \beta\hbar/2$ due to finite size effects, i.e., finite $\beta$.
}
\label{fig:corr}
\end{figure}
where ${\bar K}(i\omega)$ is obtained from ${\bar K}_{i\omega_m}$, mapping the discrete variable $i\omega_m$ into a continuous variable $i\omega$.
It should be noted that ${\bar K}(i\omega)$ is finite as $\omega \to \infty$, so that the integral is always convergent at high frequencies. This remains true also in the limiting case $\omega_c \to \infty$. Indeed, as $\omega \to \infty$, ${\bar K }(i\omega)$ diverges as $\omega^s$, which in turn leads to integrand decay at least as $\omega^{-2}$: it ensures convergence for all $s$. A possible divergence of the integral can only arise in the limit of vanishing temperature, i.e., $2\pi/(\beta\hbar) \to 0$, due to the low-frequency behavior of the denominator. Since the dissipative kernel decays as $K(\tau \to \infty) \sim \tau^{-1-s}$, the corresponding Fourier transform is such that ${\bar K}(i\omega \to 0) \sim \omega^{s}$. For $s<2$ the divergence of the denominator is therefore completely controlled by the dissipative kernel. In particular, for any $\alpha>0$, the variance $\sigma^2$: (i) converges for $s<1$, (ii) diverges logarithmically as $\log\beta$ for $s=1$, (iii) diverges as $\beta^{s-1}$ for $1<s<2$, (iv) diverges as $\beta$ for $s\ge 2$ \cite{Weiss_5th_edition}. In the last case, the divergence is the same as in the absence of dissipation. 

In conclusion, in the limit $E_J/E_C=0$, no phase transition occurs at a finite value of $\alpha$. For $s\ge1$, the phase particle is delocalized for any value of the coupling with the environment. On the other hand, in the sub-Ohmic regime, the particle is localized for any $\alpha > 0$.

\textit{Ohmic dissipation $(s=1)$: evidence of the BKT nature of the Schmid transition.---}
In this section, we prove that, defining a suitable order parameter, the Schmid transition, predicted in the Ohmic regime $s=1$, belongs to the BKT universality class for different values of $E_J/E_C$.
When $\omega_c \to \infty$ and $\Delta \tau \to 0$, the transition is expected to take place at $\alpha_c = 1$ for any $E_J/E_C>0$~\cite{Schmid, Bulgadaev}, while a slight renormalization is expected for a finite cutoff $\omega_c$ and a finite discrete time-step $\Delta \tau$. In the following, we focus our attention on the most interesting regime, i.e. $E_J/E_C \lesssim 1$, where Coulomb blockade plays a crucial role. We will show that it is possible to observe the characteristic logarithmic decay of the correlation function associated with the order parameter. 

For the purpose of defining the order parameter, we subdivide the real axis into intervals $[(2n-1)\pi,(2n+1)\pi)$ with $n \in \mathbb{Z}$, each centered around a different minimum of the cosine potential. In this way, every well can be labeled by an integer $n$. We then map the phase path $\varphi(\tau)$ onto a binary observable $S(\tau)$, such that $S(\tau)=1$ if $\varphi(\tau)$ belongs to an even potential well, and $S(\tau)=-1$ otherwise. Formally, 
\begin{equation}
     S(\tau)=(-1)^{\nint\left[\frac{\varphi(\tau)}{2\pi}\right]},
\end{equation}
where $\nint[x]$ denotes the nearest integer to $x$. The order parameter is then defined as
\begin{equation}
m^2=\frac{1}{\beta\hbar}\int\limits_0^{\beta\hbar}d\tau\,\thermal{S(\tau)S(0)}.
\end{equation}
In other words, the worldline is mapped onto a sequence of instantons and anti-instantons: the use of this order parameter allows us to focus exclusively on localization effects, without being obscured by fluctuations around potential minima. A key observation is that $m^2$ vanishes exactly in the thermodynamic limit $\beta \to +\infty$ if $\sigma^2 \to \infty$, unless the symmetry between the even and the odd minima of the potential is broken. In the latter case $m^2$ remains larger than zero for $\beta \to \infty$ and the system is in a localized phase~\footnote{We emphasize that $m^2$ may remain nonzero even in a delocalized phase if the particle preferentially occupies either even or odd minima of the potential as it happens in the case of a Josephson junction where dissipation is induced by quasi-particle tunneling~\cite{Capone}.}. 

To establish the BKT universality class of the transition, we introduce the scaling function $\Psi(\alpha,\beta)=\alpha m^2$. Within the BKT framework, and for large $\beta$, this function follows the asymptotic behavior $\Psi(\alpha_c,\beta)/\Psi_c = 1 + \frac{1}{2(\ln \beta - \ln \beta_0)}$~\cite{Minn_PRB1,Minn_PRL,Minn_PRB2}. Here, $\Psi_c$ and $\beta_0$ are fitting parameters, with $\Psi_c$ representing the jump of $\alpha m^2$ at the criticality in the thermodynamic limit. As a direct consequence of this asymptotic form, the function $G(\alpha,\beta)=\frac{1}{\Psi(\alpha,\beta)/\Psi_c-1}-2\ln\beta$ is expected to become independent of $\beta$ at very low temperatures when evaluated at criticality~\cite{Grazia_Comm_phys, Giulio_PRL, Giulio_manyspin_PRB,Giulio_spinboson_PRB}. Since no renormalization-group prediction for $\Psi_c$ is currently available, we adjusted its value until the curves became nearly independent of $\beta$ in the vicinity of the expected critical point $\alpha_c \approx 1$. This procedure provides a rough but quantitative estimate of the jump and supports the emergence of a BKT-like critical behavior in that region, as can be seen in Fig.~\ref{fig:minn}, where we show $G(\alpha, \beta)$ for different values of $E_J/E_C$, together with the corresponding estimate of the jump and of the renormalized critical point.

It is particularly interesting to observe, for large $\beta$, the long-time behavior of the correlation function $\thermal{S(\tau)S(0)}$ associated with the order parameter at the critical point obtained by the previous analysis. In fact, in the BKT framework, it is expected to behave as $\langle S(\tau)S(0)\rangle = A + \frac{B}{\ln \tau + C}$~\cite{Bhattacharjee_log_corr, Luijten_log_corr}. The corresponding fit is assumed to be reliable for $\hbar/E_C\ll\tau\ll\beta\hbar/2$, in fact, deviations are expected around $\tau \approx \beta \hbar/2$ due to finite-size effects, i.e. a finite value of $\beta$. In Fig.~\ref{fig:corr}, we show the expected logarithmic decay of correlation functions at the critical point. 
 
To provide further evidence for the universality class, we also employ a different scaling argument based on the estimate of the jump of the previous approach. Given $\Psi_c$, we interpolate $\Psi(\alpha, \beta)$ and determine $\alpha_c(\beta)$ from the intersection. We then fit the resulting values using the functional form $\alpha_c(\beta) = \alpha_c + \frac{D}{2\ln\beta + E}$~\cite{Boninsegni}. The logarithmic behavior of $\alpha_c(\beta)$ exhibits overall agreement with the previously estimated critical point for all values considered of $E_J/E_C$~\cite{EndMatterRef}.
 
The limiting cases $E_J/E_C \to 0$ and $E_J/E_C > 1$ present computational challenges. In the former, the difficulty arises from the presence of a small but non-vanishing jump, which becomes reliably estimable only at sufficiently large values of $\beta$~\cite{EndMatterRef}. In the $E_J/E_C > 1$ regime, the issue is related to an apparent non-zero temperature localization for $\alpha \lesssim 1$. Indeed, the large potential barrier, combined with finite-temperature decoherence effects, suppresses the occurrence of phase slips--i.e., quantum tunneling--except at large imaginary times, which means sufficiently large $\beta$. Nevertheless, since the underlying mechanism of the transition remains the same for any value of $E_J/E_C$ \cite{PhysRevB_Paris, PhysRevLett_Giacomelli_Devoret}, there is no reason to conclude that the universality class changes in these limiting regimes, which are therefore also expected to belong to the BKT class.

\textit{Sub-Ohmic $(s<1)$ and super-Ohmic $(s>1)$ regimes.---}
We now investigate the sub-Ohmic ($0<s<1$) and super-Ohmic ($s>1$) regimes of Eq.~\eqref{eqn:spectral_form}. Unlike the Ohmic case, a QPT is absent here~\cite{Weiss_5th_edition, Fabrizio_subOhmic}. For any $\alpha>0$ and $E_J/E_C \geq 0$, the phase particle is always localized for $s<1$ and delocalized for $s>1$. This follows from the $K(\tau) \propto \tau^{-1-s}$ decay in Eq.~\eqref{eqn:action}: slowly decaying long-range interactions ensure localization even for small $\alpha$ ($s<1$), whereas rapid decay prevents long-range correlations at any $\alpha$ ($s>1$). Thus, $s=1$ marks a critical boundary characterized by logarithmically decaying correlations~\cite{EndMatterRef}. In this context, our Renormalization Group (RG) analysis proves that the flow equations for the couplings are
\begin{equation}
\begin{split}
    \frac{d \alpha}{dl} &= -\delta s \,\alpha,\\
    \frac{dE_J}{dl} &= \biggl(1-\frac{\bar{\alpha}_s}{ \alpha}\biggr)E_J,
\end{split}
\end{equation}
where $\delta s = s-1$ and ${\bar\alpha}_s$ is a continuous positive function of $s$ in the range $0<s<2$ such that $\lim_{s\to 1} {\bar \alpha}_s = 1$~\cite{SuppMatRef}. This result proves that $-\delta s$ acts as the scaling dimension of $\alpha$, making it a relevant (irrelevant) perturbation in the sub-Ohmic (super-Ohmic) regime. In contrast, $\alpha$ is strictly marginal in the Ohmic regime. The quantity $\delta s$ also allows us to determine an inverse temperature scale $\tilde{\beta}(\alpha, s) \propto \alpha^{1/\delta s}$ at which the system starts to become distinguishable from the Ohmic regime (specifically, for $\alpha>1$ when $\delta s>0$ and $\alpha<1$ when $\delta s<0$).

\begin{figure}[]
\centering
\includegraphics[width=\columnwidth]{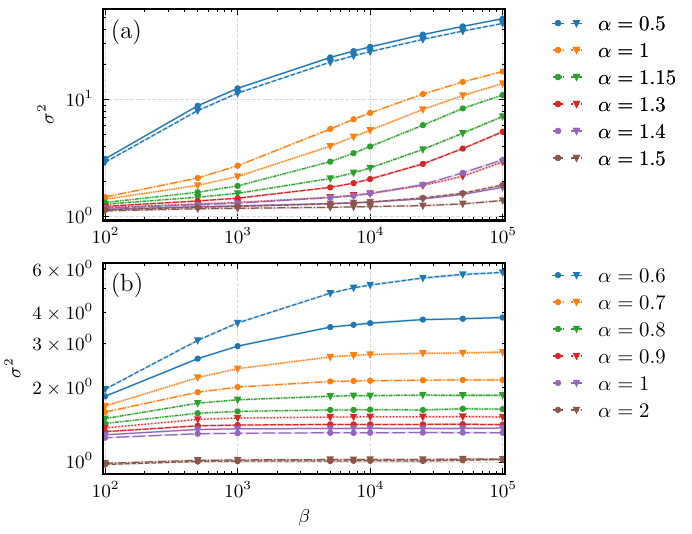}
\caption{$\sigma^2$ at $E_J/E_C = 0.5$ as a function of $\beta$ (in units of $1/E_C$) for the spectral function~\eqref{eqn:spectral_form2} at $ s=1.1$ (a) and $ s=0.9$ (b), for different values of the coupling $\alpha$. The lower cutoff is set to ${\tilde \omega}=0.5$ (circles) and ${\tilde \omega}=0.2$ (triangles).
}
\label{fig:sinf}
\end{figure}

So far, we have shown that for $E_J/E_C > 0$ in the Ohmic regime ($s = 1$), the model exhibits a QPT at a finite critical value $\alpha_c > 0$. In contrast to recent claims \cite{Altimiras:2023trq}, we argue that the nature of the QPT in this regime is only determined by the low-frequency linear behavior of the spectral function, rather than the presence or absence of a large-frequency cutoff. To further prove this, we investigated the dissipation induced by employing the spectral form~\eqref{eqn:spectral_form2}, with $\tilde{\omega}$ and $s$ close to $0$ and $1$ respectively. As specified before, the long-range behavior of the kernel depends only on the low frequency form of $J(\omega)$, for this reason $K(\tau) \sim \tau^{-1-s}$ for large $\tau$, exhibiting a deviation from the power law $\tau^{-2}$. According to the previous argument, we expect the system to be localized for any non-vanishing $\alpha$ in the range $0<s<1$, while delocalized for $s>1$. The results of the Monte Carlo simulations shown in Fig.~\ref{fig:sinf} confirm the previous scenario. On the other hand, we observe that, by lowering $\tilde{\omega}$, we need even larger values of $\beta$ to observe the non-Ohmic effects of the bath. This is reasonable since we expect to smoothly reproduce the effects of Ohmic dissipation if we take the limit $\tilde{\omega} \to 0$.

\textit{Conclusions.---}
We have shown that dissipation--driven localization of a quantum--Brownian particle in a cosine potential exhibits a BKT–like transition only for an Ohmic bath ($s=1$ in Eq.~\eqref{eqn:spectral_form}) when a finite Josephson potential is present. By introducing a binary order parameter that isolates phase slips between adjacent minima of the Josephson potential, and by combining finite--size scaling of the function $\Psi(\alpha,\beta)$, the scaling of the finite-size critical coupling $\alpha_c(\beta)$, and the long--time logarithmic decay of the correlation function $\langle S(\tau)S(0)\rangle$, we obtain a consistent picture of BKT criticality for any value of $E_J/E_C>0$. In contrast, in the limit $E_J/E_C=0$ no phase transition is observed. Using the general spectral form of Eq.~\eqref{eqn:spectral_form2}, we further demonstrate that the existence and nature of the transition depend solely on the low--frequency behavior of the bath spectral function $J(\omega)$, while high--frequency details do not affect the universal properties. In conclusion, while the non-Ohmic regime ($s\neq 1$) is essentially insensitive to the presence of the cosine potential, the ratio $E_J/E_C$ dramatically influences the low temperature physics in the Ohmic case, where the interplay between the periodic potential and linear low-frequency dissipation gives rise to the BKT phenomenology described above. Then the QPT turns out to be extremely fragile. Our results establish a unified picture in which critical behavior emerges only when both a periodic potential and an Ohmic bath are present at the same time, and it is entirely governed by the infrared structure of the environment. 

\textit{Acknowledgments.---}
G.D.F., R.F., C.A.P, and A.d.C. acknowledge financial support from PNRR MUR Project No. PE0000023-NQSTI. C.A.P. acknowledges funding from IQARO (Spin-orbitronic Quantum Bits in Reconfigurable 2DOxides) project of the European Union’s Horizon Europe research and innovation programme under grant agreement n. 101115190. R.F. acknowledges funding from the European Research Council (ERC), Grant agreement No. 101053159 – RAVE. N.N. was supported by JSPS KAKENHI Grant Numbers 24H00197, 24H02231 and 24K00583. N.N. was supported by the RIKEN TRIP initiative.

\bibliography{paper}

\begin{thebibliography}{53}%
\makeatletter
\providecommand \@ifxundefined [1]{%
 \@ifx{#1\undefined}
}%
\providecommand \@ifnum [1]{%
 \ifnum #1\expandafter \@firstoftwo
 \else \expandafter \@secondoftwo
 \fi
}%
\providecommand \@ifx [1]{%
 \ifx #1\expandafter \@firstoftwo
 \else \expandafter \@secondoftwo
 \fi
}%
\providecommand \natexlab [1]{#1}%
\providecommand \enquote  [1]{``#1''}%
\providecommand \bibnamefont  [1]{#1}%
\providecommand \bibfnamefont [1]{#1}%
\providecommand \citenamefont [1]{#1}%
\providecommand \href@noop [0]{\@secondoftwo}%
\providecommand \href [0]{\begingroup \@sanitize@url \@href}%
\providecommand \@href[1]{\@@startlink{#1}\@@href}%
\providecommand \@@href[1]{\endgroup#1\@@endlink}%
\providecommand \@sanitize@url [0]{\catcode `\\12\catcode `\$12\catcode `\&12\catcode `\#12\catcode `\^12\catcode `\_12\catcode `\%12\relax}%
\providecommand \@@startlink[1]{}%
\providecommand \@@endlink[0]{}%
\providecommand \url  [0]{\begingroup\@sanitize@url \@url }%
\providecommand \@url [1]{\endgroup\@href {#1}{\urlprefix }}%
\providecommand \urlprefix  [0]{URL }%
\providecommand \Eprint [0]{\href }%
\providecommand \doibase [0]{http://dx.doi.org/}%
\providecommand \selectlanguage [0]{\@gobble}%
\providecommand \bibinfo  [0]{\@secondoftwo}%
\providecommand \bibfield  [0]{\@secondoftwo}%
\providecommand \translation [1]{[#1]}%
\providecommand \BibitemOpen [0]{}%
\providecommand \bibitemStop [0]{}%
\providecommand \bibitemNoStop [0]{.\EOS\space}%
\providecommand \EOS [0]{\spacefactor3000\relax}%
\providecommand \BibitemShut  [1]{\csname bibitem#1\endcsname}%
\let\auto@bib@innerbib\@empty
%</preamble>
\bibitem [{\citenamefont {Dowling}\ and\ \citenamefont {Milburn}(2003)}]{Dowling_rev}%
  \BibitemOpen
  \bibfield  {author} {\bibinfo {author} {\bibfnamefont {J.~P.}\ \bibnamefont {Dowling}}\ and\ \bibinfo {author} {\bibfnamefont {G.~J.}\ \bibnamefont {Milburn}},\ }\href {\doibase 10.1098/rsta.2003.1227} {\bibfield  {journal} {\bibinfo  {journal} {Phil. Trans. Roy. Soc. Lond. A}\ }\textbf {\bibinfo {volume} {361}},\ \bibinfo {pages} {1655} (\bibinfo {year} {2003})}\BibitemShut {NoStop}%
\bibitem [{\citenamefont {Ac{\'\i}n}\ \emph {et~al.}(2018)\citenamefont {Ac{\'\i}n} \emph {et~al.}}]{Acin_rev}%
  \BibitemOpen
  \bibfield  {author} {\bibinfo {author} {\bibfnamefont {A.}~\bibnamefont {Ac{\'\i}n}} \emph {et~al.},\ }\href {\doibase 10.1088/1367-2630/aad1ea} {\bibfield  {journal} {\bibinfo  {journal} {New J. Phys.}\ }\textbf {\bibinfo {volume} {20}},\ \bibinfo {pages} {080201} (\bibinfo {year} {2018})}\BibitemShut {NoStop}%
\bibitem [{\citenamefont {Zurek}(2003)}]{Zurek_rev}%
  \BibitemOpen
  \bibfield  {author} {\bibinfo {author} {\bibfnamefont {W.~H.}\ \bibnamefont {Zurek}},\ }\href {\doibase 10.1103/RevModPhys.75.715} {\bibfield  {journal} {\bibinfo  {journal} {Rev. Mod. Phys.}\ }\textbf {\bibinfo {volume} {75}},\ \bibinfo {pages} {715} (\bibinfo {year} {2003})}\BibitemShut {NoStop}%
\bibitem [{\citenamefont {Breuer}\ and\ \citenamefont {Petruccione}(2002)}]{breuer_petruccione_book}%
  \BibitemOpen
  \bibfield  {author} {\bibinfo {author} {\bibfnamefont {H.-P.}\ \bibnamefont {Breuer}}\ and\ \bibinfo {author} {\bibfnamefont {F.}~\bibnamefont {Petruccione}},\ }\href@noop {} {\emph {\bibinfo {title} {The theory of open quantum systems}}}\ (\bibinfo  {publisher} {OUP Oxford},\ \bibinfo {year} {2002})\BibitemShut {NoStop}%
\bibitem [{\citenamefont {Caldeira}\ and\ \citenamefont {Leggett}(1983)}]{Caldeira_Leggett_1}%
  \BibitemOpen
  \bibfield  {author} {\bibinfo {author} {\bibfnamefont {A.~O.}\ \bibnamefont {Caldeira}}\ and\ \bibinfo {author} {\bibfnamefont {A.~J.}\ \bibnamefont {Leggett}},\ }\href {\doibase 10.1016/0378-4371(83)90013-4} {\bibfield  {journal} {\bibinfo  {journal} {Physica A}\ }\textbf {\bibinfo {volume} {121}},\ \bibinfo {pages} {587} (\bibinfo {year} {1983})}\BibitemShut {NoStop}%
\bibitem [{\citenamefont {Grabert}\ \emph {et~al.}(1988)\citenamefont {Grabert}, \citenamefont {Schramm},\ and\ \citenamefont {Ingold}}]{Grabert}%
  \BibitemOpen
  \bibfield  {author} {\bibinfo {author} {\bibfnamefont {H.}~\bibnamefont {Grabert}}, \bibinfo {author} {\bibfnamefont {P.}~\bibnamefont {Schramm}}, \ and\ \bibinfo {author} {\bibfnamefont {G.-L.}\ \bibnamefont {Ingold}},\ }\href {\doibase https://doi.org/10.1016/0370-1573(88)90023-3} {\bibfield  {journal} {\bibinfo  {journal} {Physics Reports}\ }\textbf {\bibinfo {volume} {168}},\ \bibinfo {pages} {115} (\bibinfo {year} {1988})}\BibitemShut {NoStop}%
\bibitem [{\citenamefont {Caldeira}\ and\ \citenamefont {Leggett}(1981)}]{Caldeira_Leggett_2}%
  \BibitemOpen
  \bibfield  {author} {\bibinfo {author} {\bibfnamefont {A.~O.}\ \bibnamefont {Caldeira}}\ and\ \bibinfo {author} {\bibfnamefont {A.~J.}\ \bibnamefont {Leggett}},\ }\href {\doibase 10.1103/PhysRevLett.46.211} {\bibfield  {journal} {\bibinfo  {journal} {Phys. Rev. Lett.}\ }\textbf {\bibinfo {volume} {46}},\ \bibinfo {pages} {211} (\bibinfo {year} {1981})}\BibitemShut {NoStop}%
\bibitem [{\citenamefont {Chakravarty}(1982)}]{Chakravarty}%
  \BibitemOpen
  \bibfield  {author} {\bibinfo {author} {\bibfnamefont {S.}~\bibnamefont {Chakravarty}},\ }\href {\doibase 10.1103/PhysRevLett.49.681} {\bibfield  {journal} {\bibinfo  {journal} {Phys. Rev. Lett.}\ }\textbf {\bibinfo {volume} {49}},\ \bibinfo {pages} {681} (\bibinfo {year} {1982})}\BibitemShut {NoStop}%
\bibitem [{\citenamefont {Sachdev}(2011)}]{Sachdev_book}%
  \BibitemOpen
  \bibfield  {author} {\bibinfo {author} {\bibfnamefont {S.}~\bibnamefont {Sachdev}},\ }\href {\doibase 10.1017/cbo9780511973765} {\emph {\bibinfo {title} {{Quantum Phase Transitions}}}}\ (\bibinfo  {publisher} {Cambridge University Press},\ \bibinfo {year} {2011})\BibitemShut {NoStop}%
\bibitem [{\citenamefont {Schmid}(1983)}]{Schmid}%
  \BibitemOpen
  \bibfield  {author} {\bibinfo {author} {\bibfnamefont {A.}~\bibnamefont {Schmid}},\ }\href {\doibase 10.1103/PhysRevLett.51.1506} {\bibfield  {journal} {\bibinfo  {journal} {Phys. Rev. Lett.}\ }\textbf {\bibinfo {volume} {51}},\ \bibinfo {pages} {1506} (\bibinfo {year} {1983})}\BibitemShut {NoStop}%
\bibitem [{\citenamefont {Bulgadaev}(1984)}]{Bulgadaev}%
  \BibitemOpen
  \bibfield  {author} {\bibinfo {author} {\bibfnamefont {S.}~\bibnamefont {Bulgadaev}},\ }\href@noop {} {\bibfield  {journal} {\bibinfo  {journal} {ZhETF Pisma Redaktsiiu}\ } (\bibinfo {year} {1984})}\BibitemShut {NoStop}%
\bibitem [{\citenamefont {Fisher}\ and\ \citenamefont {Zwerger}(1985)}]{Fisher_Zwerger}%
  \BibitemOpen
  \bibfield  {author} {\bibinfo {author} {\bibfnamefont {M.~P.~A.}\ \bibnamefont {Fisher}}\ and\ \bibinfo {author} {\bibfnamefont {W.}~\bibnamefont {Zwerger}},\ }\href {\doibase 10.1103/PhysRevB.32.6190} {\bibfield  {journal} {\bibinfo  {journal} {Phys. Rev. B}\ }\textbf {\bibinfo {volume} {32}},\ \bibinfo {pages} {6190} (\bibinfo {year} {1985})}\BibitemShut {NoStop}%
\bibitem [{\citenamefont {Weiss}(2021)}]{Weiss_5th_edition}%
  \BibitemOpen
  \bibfield  {author} {\bibinfo {author} {\bibfnamefont {U.}~\bibnamefont {Weiss}},\ }\href {\doibase 10.1142/12402} {\emph {\bibinfo {title} {{Quantum Dissipative Systems}}}}\ (\bibinfo  {publisher} {World Scientific},\ \bibinfo {year} {2021})\BibitemShut {NoStop}%
\bibitem [{\citenamefont {Werner}\ and\ \citenamefont {Troyer}(2005{\natexlab{a}})}]{Troyer_MC}%
  \BibitemOpen
  \bibfield  {author} {\bibinfo {author} {\bibfnamefont {P.}~\bibnamefont {Werner}}\ and\ \bibinfo {author} {\bibfnamefont {M.}~\bibnamefont {Troyer}},\ }\href {\doibase 10.1143/PTPS.160.395} {\bibfield  {journal} {\bibinfo  {journal} {Prog. Theor. Phys. Supp.}\ }\textbf {\bibinfo {volume} {160}},\ \bibinfo {pages} {395} (\bibinfo {year} {2005}{\natexlab{a}})}\BibitemShut {NoStop}%
\bibitem [{\citenamefont {Werner}\ and\ \citenamefont {Troyer}(2005{\natexlab{b}})}]{Troyer_PRL}%
  \BibitemOpen
  \bibfield  {author} {\bibinfo {author} {\bibfnamefont {P.}~\bibnamefont {Werner}}\ and\ \bibinfo {author} {\bibfnamefont {M.}~\bibnamefont {Troyer}},\ }\href {\doibase 10.1103/PhysRevLett.95.060201} {\bibfield  {journal} {\bibinfo  {journal} {Phys. Rev. Lett.}\ }\textbf {\bibinfo {volume} {95}},\ \bibinfo {pages} {060201} (\bibinfo {year} {2005}{\natexlab{b}})}\BibitemShut {NoStop}%
\bibitem [{\citenamefont {Hakonen}\ and\ \citenamefont {Sonin}(2021)}]{Comment_absence_Hakonen_Sonin}%
  \BibitemOpen
  \bibfield  {author} {\bibinfo {author} {\bibfnamefont {P.~J.}\ \bibnamefont {Hakonen}}\ and\ \bibinfo {author} {\bibfnamefont {E.~B.}\ \bibnamefont {Sonin}},\ }\href {\doibase 10.1103/PhysRevX.11.018001} {\bibfield  {journal} {\bibinfo  {journal} {Phys. Rev. X}\ }\textbf {\bibinfo {volume} {11}},\ \bibinfo {pages} {018001} (\bibinfo {year} {2021})}\BibitemShut {NoStop}%
\bibitem [{\citenamefont {Remez}\ \emph {et~al.}(2024)\citenamefont {Remez}, \citenamefont {Kurilovich}, \citenamefont {Rieger},\ and\ \citenamefont {Glazman}}]{Glazman_1}%
  \BibitemOpen
  \bibfield  {author} {\bibinfo {author} {\bibfnamefont {B.}~\bibnamefont {Remez}}, \bibinfo {author} {\bibfnamefont {V.~D.}\ \bibnamefont {Kurilovich}}, \bibinfo {author} {\bibfnamefont {M.}~\bibnamefont {Rieger}}, \ and\ \bibinfo {author} {\bibfnamefont {L.~I.}\ \bibnamefont {Glazman}},\ }\href {\doibase 10.1103/PhysRevB.110.054508} {\bibfield  {journal} {\bibinfo  {journal} {Phys. Rev. B}\ }\textbf {\bibinfo {volume} {110}},\ \bibinfo {pages} {054508} (\bibinfo {year} {2024})}\BibitemShut {NoStop}%
\bibitem [{\citenamefont {Houzet}\ and\ \citenamefont {Glazman}(2020)}]{Glazman_2}%
  \BibitemOpen
  \bibfield  {author} {\bibinfo {author} {\bibfnamefont {M.}~\bibnamefont {Houzet}}\ and\ \bibinfo {author} {\bibfnamefont {L.~I.}\ \bibnamefont {Glazman}},\ }\href {\doibase 10.1103/PhysRevLett.125.267701} {\bibfield  {journal} {\bibinfo  {journal} {Phys. Rev. Lett.}\ }\textbf {\bibinfo {volume} {125}},\ \bibinfo {pages} {267701} (\bibinfo {year} {2020})}\BibitemShut {NoStop}%
\bibitem [{\citenamefont {Kuzmin}\ \emph {et~al.}(2021)\citenamefont {Kuzmin}, \citenamefont {Grabon}, \citenamefont {Mehta}, \citenamefont {Burshtein}, \citenamefont {Goldstein}, \citenamefont {Houzet}, \citenamefont {Glazman},\ and\ \citenamefont {Manucharyan}}]{Glazman_3}%
  \BibitemOpen
  \bibfield  {author} {\bibinfo {author} {\bibfnamefont {R.}~\bibnamefont {Kuzmin}}, \bibinfo {author} {\bibfnamefont {N.}~\bibnamefont {Grabon}}, \bibinfo {author} {\bibfnamefont {N.}~\bibnamefont {Mehta}}, \bibinfo {author} {\bibfnamefont {A.}~\bibnamefont {Burshtein}}, \bibinfo {author} {\bibfnamefont {M.}~\bibnamefont {Goldstein}}, \bibinfo {author} {\bibfnamefont {M.}~\bibnamefont {Houzet}}, \bibinfo {author} {\bibfnamefont {L.~I.}\ \bibnamefont {Glazman}}, \ and\ \bibinfo {author} {\bibfnamefont {V.~E.}\ \bibnamefont {Manucharyan}},\ }\href {\doibase 10.1103/PhysRevLett.126.197701} {\bibfield  {journal} {\bibinfo  {journal} {Phys. Rev. Lett.}\ }\textbf {\bibinfo {volume} {126}},\ \bibinfo {pages} {197701} (\bibinfo {year} {2021})}\BibitemShut {NoStop}%
\bibitem [{\citenamefont {Houzet}\ \emph {et~al.}(2024)\citenamefont {Houzet}, \citenamefont {Yamamoto},\ and\ \citenamefont {Glazman}}]{Glazman_4}%
  \BibitemOpen
  \bibfield  {author} {\bibinfo {author} {\bibfnamefont {M.}~\bibnamefont {Houzet}}, \bibinfo {author} {\bibfnamefont {T.}~\bibnamefont {Yamamoto}}, \ and\ \bibinfo {author} {\bibfnamefont {L.~I.}\ \bibnamefont {Glazman}},\ }\href {\doibase 10.1103/PhysRevB.109.155431} {\bibfield  {journal} {\bibinfo  {journal} {Phys. Rev. B}\ }\textbf {\bibinfo {volume} {109}},\ \bibinfo {pages} {155431} (\bibinfo {year} {2024})}\BibitemShut {NoStop}%
\bibitem [{\citenamefont {Kurilovich}\ \emph {et~al.}(2025)\citenamefont {Kurilovich}, \citenamefont {Remez},\ and\ \citenamefont {Glazman}}]{Glazman_5}%
  \BibitemOpen
  \bibfield  {author} {\bibinfo {author} {\bibfnamefont {V.~D.}\ \bibnamefont {Kurilovich}}, \bibinfo {author} {\bibfnamefont {B.}~\bibnamefont {Remez}}, \ and\ \bibinfo {author} {\bibfnamefont {L.~I.}\ \bibnamefont {Glazman}},\ }\href {\doibase 10.1038/s41467-025-56411-x} {\bibfield  {journal} {\bibinfo  {journal} {Nature Commun.}\ }\textbf {\bibinfo {volume} {16}},\ \bibinfo {pages} {1384} (\bibinfo {year} {2025})}\BibitemShut {NoStop}%
\bibitem [{\citenamefont {Yamamoto}\ \emph {et~al.}(2024)\citenamefont {Yamamoto}, \citenamefont {Glazman},\ and\ \citenamefont {Houzet}}]{Glazman_6}%
  \BibitemOpen
  \bibfield  {author} {\bibinfo {author} {\bibfnamefont {T.}~\bibnamefont {Yamamoto}}, \bibinfo {author} {\bibfnamefont {L.~I.}\ \bibnamefont {Glazman}}, \ and\ \bibinfo {author} {\bibfnamefont {M.}~\bibnamefont {Houzet}},\ }\href {\doibase 10.1103/PhysRevB.110.L060512} {\bibfield  {journal} {\bibinfo  {journal} {Phys. Rev. B}\ }\textbf {\bibinfo {volume} {110}},\ \bibinfo {pages} {L060512} (\bibinfo {year} {2024})}\BibitemShut {NoStop}%
\bibitem [{\citenamefont {Murani}\ \emph {et~al.}(2020)\citenamefont {Murani}, \citenamefont {Bourlet}, \citenamefont {le~Sueur}, \citenamefont {Portier}, \citenamefont {Altimiras}, \citenamefont {Esteve}, \citenamefont {Grabert}, \citenamefont {Stockburger}, \citenamefont {Ankerhold},\ and\ \citenamefont {Joyez}}]{Murani_absence}%
  \BibitemOpen
  \bibfield  {author} {\bibinfo {author} {\bibfnamefont {A.}~\bibnamefont {Murani}}, \bibinfo {author} {\bibfnamefont {N.}~\bibnamefont {Bourlet}}, \bibinfo {author} {\bibfnamefont {H.}~\bibnamefont {le~Sueur}}, \bibinfo {author} {\bibfnamefont {F.}~\bibnamefont {Portier}}, \bibinfo {author} {\bibfnamefont {C.}~\bibnamefont {Altimiras}}, \bibinfo {author} {\bibfnamefont {D.}~\bibnamefont {Esteve}}, \bibinfo {author} {\bibfnamefont {H.}~\bibnamefont {Grabert}}, \bibinfo {author} {\bibfnamefont {J.}~\bibnamefont {Stockburger}}, \bibinfo {author} {\bibfnamefont {J.}~\bibnamefont {Ankerhold}}, \ and\ \bibinfo {author} {\bibfnamefont {P.}~\bibnamefont {Joyez}},\ }\href {\doibase 10.1103/PhysRevX.10.021003} {\bibfield  {journal} {\bibinfo  {journal} {Phys. Rev. X}\ }\textbf {\bibinfo {volume} {10}},\ \bibinfo {pages} {021003} (\bibinfo {year} {2020})}\BibitemShut {NoStop}%
\bibitem [{\citenamefont {Murani}\ \emph {et~al.}(2021)\citenamefont {Murani}, \citenamefont {Bourlet}, \citenamefont {le~Sueur}, \citenamefont {Portier}, \citenamefont {Altimiras}, \citenamefont {Esteve}, \citenamefont {Grabert}, \citenamefont {Stockburger}, \citenamefont {Ankerhold},\ and\ \citenamefont {Joyez}}]{Murani_reply}%
  \BibitemOpen
  \bibfield  {author} {\bibinfo {author} {\bibfnamefont {A.}~\bibnamefont {Murani}}, \bibinfo {author} {\bibfnamefont {N.}~\bibnamefont {Bourlet}}, \bibinfo {author} {\bibfnamefont {H.}~\bibnamefont {le~Sueur}}, \bibinfo {author} {\bibfnamefont {F.}~\bibnamefont {Portier}}, \bibinfo {author} {\bibfnamefont {C.}~\bibnamefont {Altimiras}}, \bibinfo {author} {\bibfnamefont {D.}~\bibnamefont {Esteve}}, \bibinfo {author} {\bibfnamefont {H.}~\bibnamefont {Grabert}}, \bibinfo {author} {\bibfnamefont {J.}~\bibnamefont {Stockburger}}, \bibinfo {author} {\bibfnamefont {J.}~\bibnamefont {Ankerhold}}, \ and\ \bibinfo {author} {\bibfnamefont {P.}~\bibnamefont {Joyez}},\ }\href {\doibase 10.1103/PhysRevX.11.018002} {\bibfield  {journal} {\bibinfo  {journal} {Phys. Rev. X}\ }\textbf {\bibinfo {volume} {11}},\ \bibinfo {pages} {018002} (\bibinfo {year} {2021})}\BibitemShut {NoStop}%
\bibitem [{\citenamefont {Altimiras}\ \emph {et~al.}(2023)\citenamefont {Altimiras}, \citenamefont {Esteve}, \citenamefont {Girit}, \citenamefont {Le~Sueur},\ and\ \citenamefont {Joyez}}]{Altimiras:2023trq}%
  \BibitemOpen
  \bibfield  {author} {\bibinfo {author} {\bibfnamefont {C.}~\bibnamefont {Altimiras}}, \bibinfo {author} {\bibfnamefont {D.}~\bibnamefont {Esteve}}, \bibinfo {author} {\bibfnamefont {{\c{C}}.}~\bibnamefont {Girit}}, \bibinfo {author} {\bibfnamefont {H.}~\bibnamefont {Le~Sueur}}, \ and\ \bibinfo {author} {\bibfnamefont {P.}~\bibnamefont {Joyez}},\ }\href@noop {} {\  (\bibinfo {year} {2023})},\ \Eprint {http://arxiv.org/abs/2312.14754} {arXiv:2312.14754 [cond-mat.supr-con]} \BibitemShut {NoStop}%
\bibitem [{The Monte Carlo procedure proposed in Ref. \cite{Altimiras:2023trq}, based on a Hubbard-Stratonovich transformation and an auxiliary random field()}]{notaa}%
  \BibitemOpen
  The Monte Carlo procedure proposed in Ref. \cite{Altimiras:2023trq}, based on a Hubbard-Stratonovich transformation and an auxiliary random field,\ \href@noop {} {}\bibinfo {note} {{l}ikely suffers from prohibitively slow convergence and fails to thermalize within accessible timescales. Furthermore, the authors argue that fixing the auxiliary field $\xi$ breaks the translational invariance of the marginal probability distribution \textit{a priori}, leading to a ground-state degeneracy for any shunt resistance and the absence of a QPT. We contend, however, that adopting the joint distribution of $\xi$ and $\varphi$ would restore full translational symmetry, yielding a QPT at a critical value of the shunt resistance.}\BibitemShut {Stop}%
\bibitem [{\citenamefont {Masuki}\ \emph {et~al.}(2022)\citenamefont {Masuki}, \citenamefont {Sudo}, \citenamefont {Oshikawa},\ and\ \citenamefont {Ashida}}]{Masuki_Absence}%
  \BibitemOpen
  \bibfield  {author} {\bibinfo {author} {\bibfnamefont {K.}~\bibnamefont {Masuki}}, \bibinfo {author} {\bibfnamefont {H.}~\bibnamefont {Sudo}}, \bibinfo {author} {\bibfnamefont {M.}~\bibnamefont {Oshikawa}}, \ and\ \bibinfo {author} {\bibfnamefont {Y.}~\bibnamefont {Ashida}},\ }\href {\doibase 10.1103/PhysRevLett.129.087001} {\bibfield  {journal} {\bibinfo  {journal} {Phys. Rev. Lett.}\ }\textbf {\bibinfo {volume} {129}},\ \bibinfo {pages} {087001} (\bibinfo {year} {2022})}\BibitemShut {NoStop}%
\bibitem [{\citenamefont {Masuki}\ \emph {et~al.}(2023)\citenamefont {Masuki}, \citenamefont {Sudo}, \citenamefont {Oshikawa},\ and\ \citenamefont {Ashida}}]{Masuki_reply}%
  \BibitemOpen
  \bibfield  {author} {\bibinfo {author} {\bibfnamefont {K.}~\bibnamefont {Masuki}}, \bibinfo {author} {\bibfnamefont {H.}~\bibnamefont {Sudo}}, \bibinfo {author} {\bibfnamefont {M.}~\bibnamefont {Oshikawa}}, \ and\ \bibinfo {author} {\bibfnamefont {Y.}~\bibnamefont {Ashida}},\ }\href {\doibase 10.1103/PhysRevLett.131.199702} {\bibfield  {journal} {\bibinfo  {journal} {Phys. Rev. Lett.}\ }\textbf {\bibinfo {volume} {131}},\ \bibinfo {pages} {199702} (\bibinfo {year} {2023})}\BibitemShut {NoStop}%
\bibitem [{\citenamefont {S\'epulcre}\ \emph {et~al.}(2023)\citenamefont {S\'epulcre}, \citenamefont {Florens},\ and\ \citenamefont {Snyman}}]{Comment_absence_Sepulcre}%
  \BibitemOpen
  \bibfield  {author} {\bibinfo {author} {\bibfnamefont {T.}~\bibnamefont {S\'epulcre}}, \bibinfo {author} {\bibfnamefont {S.}~\bibnamefont {Florens}}, \ and\ \bibinfo {author} {\bibfnamefont {I.}~\bibnamefont {Snyman}},\ }\href {\doibase 10.1103/PhysRevLett.131.199701} {\bibfield  {journal} {\bibinfo  {journal} {Phys. Rev. Lett.}\ }\textbf {\bibinfo {volume} {131}},\ \bibinfo {pages} {199701} (\bibinfo {year} {2023})}\BibitemShut {NoStop}%
\bibitem [{Sup()}]{SuppMatRef}%
  \BibitemOpen
  \href@noop {} {}\bibinfo {note} {See Supplemental Material for further details on the WLMC algorithm and a RG analysis validating our numerical calculations, which includes Refs.~\cite{Troyer_MC,Troyer_PRL,Wolff_PRL,Luijten1,Luijten2,Fisher_Zwerger,Bulgadaev}.}\BibitemShut {Stop}%
\bibitem [{\citenamefont {Vool}\ and\ \citenamefont {Devoret}(2017)}]{Vool-Devoret}%
  \BibitemOpen
  \bibfield  {author} {\bibinfo {author} {\bibfnamefont {U.}~\bibnamefont {Vool}}\ and\ \bibinfo {author} {\bibfnamefont {M.~H.}\ \bibnamefont {Devoret}},\ }\href {\doibase 10.1002/cta.2359} {\bibfield  {journal} {\bibinfo  {journal} {Int. J. Circuit Theor. Appl.}\ }\textbf {\bibinfo {volume} {45}},\ \bibinfo {pages} {897} (\bibinfo {year} {2017})},\ \Eprint {http://arxiv.org/abs/1610.03438} {arXiv:1610.03438 [quant-ph]} \BibitemShut {NoStop}%
\bibitem [{\citenamefont {Sch\"{o}n}\ and\ \citenamefont {Zaikin}(1990)}]{Schon_Zaikin_review}%
  \BibitemOpen
  \bibfield  {author} {\bibinfo {author} {\bibfnamefont {G.}~\bibnamefont {Sch\"{o}n}}\ and\ \bibinfo {author} {\bibfnamefont {A.}~\bibnamefont {Zaikin}},\ }\href {\doibase 10.1016/0370-1573(90)90156-v} {\bibfield  {journal} {\bibinfo  {journal} {Phys. Rep.}\ }\textbf {\bibinfo {volume} {198}},\ \bibinfo {pages} {237–412} (\bibinfo {year} {1990})}\BibitemShut {NoStop}%
\bibitem [{\citenamefont {Likharev}\ and\ \citenamefont {Zorin}(1985)}]{Likharev}%
  \BibitemOpen
  \bibfield  {author} {\bibinfo {author} {\bibfnamefont {K.}~\bibnamefont {Likharev}}\ and\ \bibinfo {author} {\bibfnamefont {A.}~\bibnamefont {Zorin}},\ }\href {\doibase https://doi.org/10.1007/BF00683782} {\bibfield  {journal} {\bibinfo  {journal} {J. Low. Temp. Phys.}\ }\textbf {\bibinfo {volume} {59}},\ \bibinfo {pages} {347–382} (\bibinfo {year} {1985})}\BibitemShut {NoStop}%
\bibitem [{\citenamefont {Averin}\ \emph {et~al.}(1985)\citenamefont {Averin}, \citenamefont {Zorin},\ and\ \citenamefont {Likharev}}]{Averin}%
  \BibitemOpen
  \bibfield  {author} {\bibinfo {author} {\bibfnamefont {D.}~\bibnamefont {Averin}}, \bibinfo {author} {\bibfnamefont {A.}~\bibnamefont {Zorin}}, \ and\ \bibinfo {author} {\bibfnamefont {K.}~\bibnamefont {Likharev}},\ }\href@noop {} {\bibfield  {journal} {\bibinfo  {journal} {Sov. Phys. - JETP}\ }\textbf {\bibinfo {volume} {61}},\ \bibinfo {pages} {407–413} (\bibinfo {year} {1985})}\BibitemShut {NoStop}%
\bibitem [{End()}]{EndMatterRef}%
  \BibitemOpen
  \href@noop {} {}\bibinfo {note} {See End Matter. It contains more details on the case of vanishing cosine potential, further evidences of the BKT nature of the Schmid transition, and the results of the proposed WLMC approach in the sub-Ohmic and super-Ohmic regimes.}\BibitemShut {Stop}%
\bibitem [{Note1()}]{Note1}%
  \BibitemOpen
  \bibinfo {note} {We emphasize that $m^2$ may remain nonzero even in a delocalized phase if the particle preferentially occupies either even or odd minima of the potential as it happens in the case of a Josephson junction where dissipation is induced by quasi-particle tunneling~\cite {Capone}.}\BibitemShut {Stop}%
\bibitem [{\citenamefont {Minnhagen}(1985{\natexlab{a}})}]{Minn_PRB1}%
  \BibitemOpen
  \bibfield  {author} {\bibinfo {author} {\bibfnamefont {P.}~\bibnamefont {Minnhagen}},\ }\href {\doibase 10.1103/PhysRevB.32.3088} {\bibfield  {journal} {\bibinfo  {journal} {Phys. Rev. B}\ }\textbf {\bibinfo {volume} {32}},\ \bibinfo {pages} {3088} (\bibinfo {year} {1985}{\natexlab{a}})}\BibitemShut {NoStop}%
\bibitem [{\citenamefont {Minnhagen}(1985{\natexlab{b}})}]{Minn_PRL}%
  \BibitemOpen
  \bibfield  {author} {\bibinfo {author} {\bibfnamefont {P.}~\bibnamefont {Minnhagen}},\ }\href {\doibase 10.1103/PhysRevLett.54.2351} {\bibfield  {journal} {\bibinfo  {journal} {Phys. Rev. Lett.}\ }\textbf {\bibinfo {volume} {54}},\ \bibinfo {pages} {2351} (\bibinfo {year} {1985}{\natexlab{b}})}\BibitemShut {NoStop}%
\bibitem [{\citenamefont {Weber}\ and\ \citenamefont {Minnhagen}(1988)}]{Minn_PRB2}%
  \BibitemOpen
  \bibfield  {author} {\bibinfo {author} {\bibfnamefont {H.}~\bibnamefont {Weber}}\ and\ \bibinfo {author} {\bibfnamefont {P.}~\bibnamefont {Minnhagen}},\ }\href {\doibase 10.1103/PhysRevB.37.5986} {\bibfield  {journal} {\bibinfo  {journal} {Phys. Rev. B}\ }\textbf {\bibinfo {volume} {37}},\ \bibinfo {pages} {5986} (\bibinfo {year} {1988})}\BibitemShut {NoStop}%
\bibitem [{\citenamefont {Di~Bello}\ \emph {et~al.}(2024)\citenamefont {Di~Bello}, \citenamefont {Ponticelli}, \citenamefont {Pavan}, \citenamefont {Cataudella}, \citenamefont {De~Filippis}, \citenamefont {de~Candia},\ and\ \citenamefont {Perroni}}]{Grazia_Comm_phys}%
  \BibitemOpen
  \bibfield  {author} {\bibinfo {author} {\bibfnamefont {G.}~\bibnamefont {Di~Bello}}, \bibinfo {author} {\bibfnamefont {A.}~\bibnamefont {Ponticelli}}, \bibinfo {author} {\bibfnamefont {F.}~\bibnamefont {Pavan}}, \bibinfo {author} {\bibfnamefont {V.}~\bibnamefont {Cataudella}}, \bibinfo {author} {\bibfnamefont {G.}~\bibnamefont {De~Filippis}}, \bibinfo {author} {\bibfnamefont {A.}~\bibnamefont {de~Candia}}, \ and\ \bibinfo {author} {\bibfnamefont {C.~A.}\ \bibnamefont {Perroni}},\ }\href {\doibase 10.1038/s42005-024-01855-8} {\bibfield  {journal} {\bibinfo  {journal} {Commun. Phys.}\ }\textbf {\bibinfo {volume} {7}},\ \bibinfo {pages} {364} (\bibinfo {year} {2024})}\BibitemShut {NoStop}%
\bibitem [{\citenamefont {De~Filippis}\ \emph {et~al.}(2023)\citenamefont {De~Filippis}, \citenamefont {de~Candia}, \citenamefont {Di~Bello}, \citenamefont {Perroni}, \citenamefont {Cangemi}, \citenamefont {Nocera}, \citenamefont {Sassetti}, \citenamefont {Fazio},\ and\ \citenamefont {Cataudella}}]{Giulio_PRL}%
  \BibitemOpen
  \bibfield  {author} {\bibinfo {author} {\bibfnamefont {G.}~\bibnamefont {De~Filippis}}, \bibinfo {author} {\bibfnamefont {A.}~\bibnamefont {de~Candia}}, \bibinfo {author} {\bibfnamefont {G.}~\bibnamefont {Di~Bello}}, \bibinfo {author} {\bibfnamefont {C.~A.}\ \bibnamefont {Perroni}}, \bibinfo {author} {\bibfnamefont {L.~M.}\ \bibnamefont {Cangemi}}, \bibinfo {author} {\bibfnamefont {A.}~\bibnamefont {Nocera}}, \bibinfo {author} {\bibfnamefont {M.}~\bibnamefont {Sassetti}}, \bibinfo {author} {\bibfnamefont {R.}~\bibnamefont {Fazio}}, \ and\ \bibinfo {author} {\bibfnamefont {V.}~\bibnamefont {Cataudella}},\ }\href {\doibase 10.1103/PhysRevLett.130.210404} {\bibfield  {journal} {\bibinfo  {journal} {Phys. Rev. Lett.}\ }\textbf {\bibinfo {volume} {130}},\ \bibinfo {pages} {210404} (\bibinfo {year} {2023})}\BibitemShut {NoStop}%
\bibitem [{\citenamefont {De~Filippis}\ \emph {et~al.}(2021)\citenamefont {De~Filippis}, \citenamefont {de~Candia}, \citenamefont {Mishchenko}, \citenamefont {Cangemi}, \citenamefont {Nocera}, \citenamefont {Mishchenko}, \citenamefont {Sassetti}, \citenamefont {Fazio}, \citenamefont {Nagaosa},\ and\ \citenamefont {Cataudella}}]{Giulio_manyspin_PRB}%
  \BibitemOpen
  \bibfield  {author} {\bibinfo {author} {\bibfnamefont {G.}~\bibnamefont {De~Filippis}}, \bibinfo {author} {\bibfnamefont {A.}~\bibnamefont {de~Candia}}, \bibinfo {author} {\bibfnamefont {A.~S.}\ \bibnamefont {Mishchenko}}, \bibinfo {author} {\bibfnamefont {L.~M.}\ \bibnamefont {Cangemi}}, \bibinfo {author} {\bibfnamefont {A.}~\bibnamefont {Nocera}}, \bibinfo {author} {\bibfnamefont {P.~A.}\ \bibnamefont {Mishchenko}}, \bibinfo {author} {\bibfnamefont {M.}~\bibnamefont {Sassetti}}, \bibinfo {author} {\bibfnamefont {R.}~\bibnamefont {Fazio}}, \bibinfo {author} {\bibfnamefont {N.}~\bibnamefont {Nagaosa}}, \ and\ \bibinfo {author} {\bibfnamefont {V.}~\bibnamefont {Cataudella}},\ }\href {\doibase 10.1103/PhysRevB.104.L060410} {\bibfield  {journal} {\bibinfo  {journal} {Phys. Rev. B}\ }\textbf {\bibinfo {volume} {104}},\ \bibinfo {pages} {L060410} (\bibinfo {year} {2021})}\BibitemShut {NoStop}%
\bibitem [{\citenamefont {De~Filippis}\ \emph {et~al.}(2020)\citenamefont {De~Filippis}, \citenamefont {de~Candia}, \citenamefont {Cangemi}, \citenamefont {Sassetti}, \citenamefont {Fazio},\ and\ \citenamefont {Cataudella}}]{Giulio_spinboson_PRB}%
  \BibitemOpen
  \bibfield  {author} {\bibinfo {author} {\bibfnamefont {G.}~\bibnamefont {De~Filippis}}, \bibinfo {author} {\bibfnamefont {A.}~\bibnamefont {de~Candia}}, \bibinfo {author} {\bibfnamefont {L.~M.}\ \bibnamefont {Cangemi}}, \bibinfo {author} {\bibfnamefont {M.}~\bibnamefont {Sassetti}}, \bibinfo {author} {\bibfnamefont {R.}~\bibnamefont {Fazio}}, \ and\ \bibinfo {author} {\bibfnamefont {V.}~\bibnamefont {Cataudella}},\ }\href {\doibase 10.1103/PhysRevB.101.180408} {\bibfield  {journal} {\bibinfo  {journal} {Phys. Rev. B}\ }\textbf {\bibinfo {volume} {101}},\ \bibinfo {pages} {180408} (\bibinfo {year} {2020})}\BibitemShut {NoStop}%
\bibitem [{\citenamefont {Bhattacharjee}\ \emph {et~al.}(1981)\citenamefont {Bhattacharjee}, \citenamefont {Chakravarty}, \citenamefont {Richardson},\ and\ \citenamefont {Scalapino}}]{Bhattacharjee_log_corr}%
  \BibitemOpen
  \bibfield  {author} {\bibinfo {author} {\bibfnamefont {J.}~\bibnamefont {Bhattacharjee}}, \bibinfo {author} {\bibfnamefont {S.}~\bibnamefont {Chakravarty}}, \bibinfo {author} {\bibfnamefont {J.~L.}\ \bibnamefont {Richardson}}, \ and\ \bibinfo {author} {\bibfnamefont {D.~J.}\ \bibnamefont {Scalapino}},\ }\href {\doibase 10.1103/PhysRevB.24.3862} {\bibfield  {journal} {\bibinfo  {journal} {Phys. Rev. B}\ }\textbf {\bibinfo {volume} {24}},\ \bibinfo {pages} {3862} (\bibinfo {year} {1981})}\BibitemShut {NoStop}%
\bibitem [{\citenamefont {Luijten}\ and\ \citenamefont {Me\ss{}ingfeld}(2001)}]{Luijten_log_corr}%
  \BibitemOpen
  \bibfield  {author} {\bibinfo {author} {\bibfnamefont {E.}~\bibnamefont {Luijten}}\ and\ \bibinfo {author} {\bibfnamefont {H.}~\bibnamefont {Me\ss{}ingfeld}},\ }\href {\doibase 10.1103/PhysRevLett.86.5305} {\bibfield  {journal} {\bibinfo  {journal} {Phys. Rev. Lett.}\ }\textbf {\bibinfo {volume} {86}},\ \bibinfo {pages} {5305} (\bibinfo {year} {2001})}\BibitemShut {NoStop}%
\bibitem [{\citenamefont {Nguyen}\ and\ \citenamefont {Boninsegni}(2021)}]{Boninsegni}%
  \BibitemOpen
  \bibfield  {author} {\bibinfo {author} {\bibfnamefont {P.~H.}\ \bibnamefont {Nguyen}}\ and\ \bibinfo {author} {\bibfnamefont {M.}~\bibnamefont {Boninsegni}},\ }\href {\doibase 10.3390/app11114931} {\bibfield  {journal} {\bibinfo  {journal} {Appl. Sci.}\ }\textbf {\bibinfo {volume} {11}},\ \bibinfo {pages} {4931} (\bibinfo {year} {2021})}\BibitemShut {NoStop}%
\bibitem [{\citenamefont {Paris}\ \emph {et~al.}(2025)\citenamefont {Paris}, \citenamefont {Giacomelli}, \citenamefont {Daviet}, \citenamefont {Ciuti}, \citenamefont {Dupuis},\ and\ \citenamefont {Mora}}]{PhysRevB_Paris}%
  \BibitemOpen
  \bibfield  {author} {\bibinfo {author} {\bibfnamefont {N.}~\bibnamefont {Paris}}, \bibinfo {author} {\bibfnamefont {L.}~\bibnamefont {Giacomelli}}, \bibinfo {author} {\bibfnamefont {R.}~\bibnamefont {Daviet}}, \bibinfo {author} {\bibfnamefont {C.}~\bibnamefont {Ciuti}}, \bibinfo {author} {\bibfnamefont {N.}~\bibnamefont {Dupuis}}, \ and\ \bibinfo {author} {\bibfnamefont {C.}~\bibnamefont {Mora}},\ }\href {\doibase 10.1103/PhysRevB.111.064509} {\bibfield  {journal} {\bibinfo  {journal} {Phys. Rev. B}\ }\textbf {\bibinfo {volume} {111}},\ \bibinfo {pages} {064509} (\bibinfo {year} {2025})}\BibitemShut {NoStop}%
\bibitem [{\citenamefont {Giacomelli}\ \emph {et~al.}(2026)\citenamefont {Giacomelli}, \citenamefont {Devoret},\ and\ \citenamefont {Ciuti}}]{PhysRevLett_Giacomelli_Devoret}%
  \BibitemOpen
  \bibfield  {author} {\bibinfo {author} {\bibfnamefont {L.}~\bibnamefont {Giacomelli}}, \bibinfo {author} {\bibfnamefont {M.~H.}\ \bibnamefont {Devoret}}, \ and\ \bibinfo {author} {\bibfnamefont {C.}~\bibnamefont {Ciuti}},\ }\href {\doibase 10.1103/1k3s-m3b9} {\bibfield  {journal} {\bibinfo  {journal} {Phys. Rev. Lett.}\ ,\ } (\bibinfo {year} {2026})}\BibitemShut {NoStop}%
\bibitem [{\citenamefont {Fabrizio}\ \emph {et~al.}(1994)\citenamefont {Fabrizio}, \citenamefont {Gogolin},\ and\ \citenamefont {Scheidl}}]{Fabrizio_subOhmic}%
  \BibitemOpen
  \bibfield  {author} {\bibinfo {author} {\bibfnamefont {M.}~\bibnamefont {Fabrizio}}, \bibinfo {author} {\bibfnamefont {A.~O.}\ \bibnamefont {Gogolin}}, \ and\ \bibinfo {author} {\bibfnamefont {S.}~\bibnamefont {Scheidl}},\ }\href {\doibase 10.1103/PhysRevLett.72.2235} {\bibfield  {journal} {\bibinfo  {journal} {Phys. Rev. Lett.}\ }\textbf {\bibinfo {volume} {72}},\ \bibinfo {pages} {2235} (\bibinfo {year} {1994})}\BibitemShut {NoStop}%
\bibitem [{\citenamefont {Wolff}(1989)}]{Wolff_PRL}%
  \BibitemOpen
  \bibfield  {author} {\bibinfo {author} {\bibfnamefont {U.}~\bibnamefont {Wolff}},\ }\href {\doibase 10.1103/PhysRevLett.62.361} {\bibfield  {journal} {\bibinfo  {journal} {Phys. Rev. Lett.}\ }\textbf {\bibinfo {volume} {62}},\ \bibinfo {pages} {361} (\bibinfo {year} {1989})}\BibitemShut {NoStop}%
\bibitem [{\citenamefont {Luijten}\ and\ \citenamefont {Bl\"{o}te}(1995)}]{Luijten1}%
  \BibitemOpen
  \bibfield  {author} {\bibinfo {author} {\bibfnamefont {E.}~\bibnamefont {Luijten}}\ and\ \bibinfo {author} {\bibfnamefont {H.~W.}\ \bibnamefont {Bl\"{o}te}},\ }\href {\doibase 10.1142/S0129183195000265} {\bibfield  {journal} {\bibinfo  {journal} {Int. J. Mod. Phys. C}\ }\textbf {\bibinfo {volume} {06}},\ \bibinfo {pages} {359} (\bibinfo {year} {1995})}\BibitemShut {NoStop}%
\bibitem [{\citenamefont {Luijten}(2000)}]{Luijten2}%
  \BibitemOpen
  \bibfield  {author} {\bibinfo {author} {\bibfnamefont {E.}~\bibnamefont {Luijten}},\ }in\ \href@noop {} {\emph {\bibinfo {booktitle} {Computer Simulation Studies in Condensed-Matter Physics XII}}},\ \bibinfo {editor} {edited by\ \bibinfo {editor} {\bibfnamefont {D.~P.}\ \bibnamefont {Landau}}, \bibinfo {editor} {\bibfnamefont {S.~P.}\ \bibnamefont {Lewis}}, \ and\ \bibinfo {editor} {\bibfnamefont {H.-B.}\ \bibnamefont {Sch{\"u}ttler}}}\ (\bibinfo  {publisher} {Springer Berlin Heidelberg},\ \bibinfo {address} {Berlin, Heidelberg},\ \bibinfo {year} {2000})\ pp.\ \bibinfo {pages} {86--99}\BibitemShut {NoStop}%
\bibitem [{\citenamefont {Capone}\ \emph {et~al.}(2025)\citenamefont {Capone}, \citenamefont {de~Candia}, \citenamefont {Cataudella}, \citenamefont {Nagaosa}, \citenamefont {Perroni},\ and\ \citenamefont {De~Filippis}}]{Capone}%
  \BibitemOpen
  \bibfield  {author} {\bibinfo {author} {\bibfnamefont {F.~G.}\ \bibnamefont {Capone}}, \bibinfo {author} {\bibfnamefont {A.}~\bibnamefont {de~Candia}}, \bibinfo {author} {\bibfnamefont {V.}~\bibnamefont {Cataudella}}, \bibinfo {author} {\bibfnamefont {N.}~\bibnamefont {Nagaosa}}, \bibinfo {author} {\bibfnamefont {C.~A.}\ \bibnamefont {Perroni}}, \ and\ \bibinfo {author} {\bibfnamefont {G.}~\bibnamefont {De~Filippis}},\ }\href@noop {} {\  (\bibinfo {year} {2025})},\ \Eprint {http://arxiv.org/abs/2504.00258} {arXiv:2504.00258 [cond-mat.supr-con]} \BibitemShut {NoStop}%
\end{thebibliography}%

\clearpage

\appendix

{\Large \bf End Matter}
\section{Appendix A: Vanishing cosine potential. Analytic results}
In this section, we discuss in more detail the limiting case of a vanishing cosine potential ($E_J = 0$), introduced in the main text. Defining the Matsubara Fourier transform of the field as $\tilde{\varphi}_{i\omega_n} = \int_0^{\beta\hbar} d\tau\, \varphi(\tau) e^{-i\omega_n\tau}$, where $\omega_n = \frac{2n\pi}{\beta\hbar}$, and $n \in \mathbb{Z}$, the functional integral on the right-hand side of \eqref{eqn:partition_path} can be expressed as an integral over the real and imaginary parts of the Fourier coefficients. Specifically, in this representation, the action \eqref{eqn:action} takes the form
\begin{equation}\begin{split}{\label{eqn:fourier_action}}
     &{\cal S}[{\tilde\varphi}_{i\omega_n}] = \frac{2}{\beta \hbar} \sum_{m=1}^{\infty}A_{i\omega_m}|{\tilde\varphi}_{i\omega_m}|^2,
\end{split}
\end{equation}
where $A_{i\omega_m} = \frac{\hbar^2}{4E_C}\omega_m^2+2{\tilde K}_{i\omega_l=0}-2{\tilde K}_{i\omega_m}$.
From the Fourier decomposition, it is straightforward to prove that $\langle \varphi^2\rangle = \frac{1}{(\beta \hbar)^2}\sum_{m=-\infty}^{\infty}\langle|\tilde{\varphi}_{i\omega_m}|^2\rangle$ and $\langle\bar \varphi ^2\rangle = \frac{1}{(\beta \hbar)^2} \langle |\tilde{\varphi}_{i\omega_m = 0}|^2\rangle$, leading to
\begin{equation}\label{eqn:varianza_1}
    \sigma^2 = \frac{2}{(\beta\hbar)^2} \sum_{m=1}^{\infty} \langle |\tilde{\varphi}_{i\omega_m}|^2 \rangle,
\end{equation}
where we used the property $\varphi_{i\omega_m} = \varphi^*_{-i\omega_m}$. Using Eq. \eqref{eqn:fourier_action}, standard gaussian results provide
\begin{equation}\label{eqn:valori_medi}
    \langle |\tilde{\varphi}_{i\omega_m}|^2 \rangle = \frac{\beta \hbar}{2} \frac{\hbar}{A_{i\omega_m}}.
\end{equation}
Replacing Eq.\eqref{eqn:valori_medi} in Eq.\eqref{eqn:varianza_1} yields Eq.\eqref{eqn:variance_Ej0} of the main text.

\section{Appendix B: BKT scaling}
\begin{figure}[h]
\centering
\includegraphics{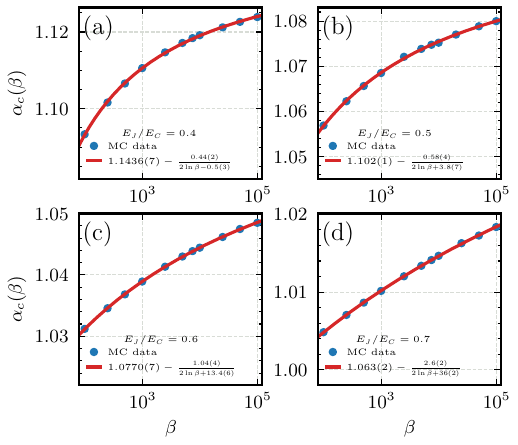}
\caption{Critical coupling at finite size (blue  dots) and corresponding logarithmic fit $\alpha_c + \frac{D}{2\ln\beta + E}$ (red lines) as functions of $\beta$ (in units of $1/E_C$) for $E_J/E_C = 0.4$ (a), $E_J/E_C = 0.5$ (b), $E_J/E_C = 0.6$ (c), $E_J/E_C = 0.7$ (d). The fitted $\alpha_c$ values are in good agreement with the Schmid critical coupling reported in the literature and calculated in the main text.
}
\label{fig:boninsegni}
\end{figure}

\begin{figure}[h]
\centering
\includegraphics{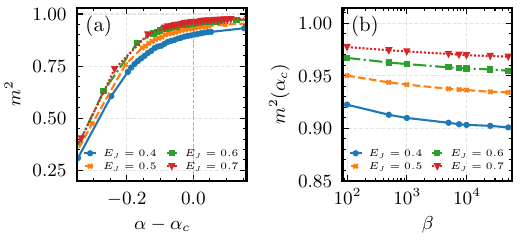}
\caption{(a) Order parameter $m^2$ as a function of $\alpha-\alpha_c$ for different values of $E_J$ (in units of $E_C$), computed at $\beta E_C = 50000$. The values of $\alpha_c$ are determined as discussed in the main text. Note the suppression of $m^2$ near the transition as $E_J/E_C$ decreases. (b) Order parameter $m^2$ as a function of $\beta$ (in units of $1/E_C$), evaluated exactly at the critical coupling $\alpha_c$. The jump in the order parameter corresponds to the asymptotic limit $\beta \to \infty$. These data confirm the reduction of the discontinuity for smaller $E_J/E_C$ values.
}
\label{fig:jump}
\end{figure}

In this section, we provide further evidence of the BKT universality class, employing a different scaling argument based on the estimate of the jump $\Psi_c$ determined by the approach introduced in the main text. Once interpolating $\Psi(\alpha, \beta)$, we determine the finite $\beta$ critical coupling solving the equation $\Psi (\alpha_c(\beta), \beta) = \Psi_c$. We then fit the resulting $\alpha_c(\beta)$ values using the functional form $\alpha_c(\beta) = \alpha_c + \frac{D}{2\ln\beta + E}$~\cite{Boninsegni}. As shown in Fig.~\ref{fig:boninsegni}, the logarithmic behavior of $\alpha_c(\beta)$ exhibits overall agreement with the critical point estimated in the main text for all values considered of $E_J/E_C$.
In addition, we show that the jump in the order parameter $m^2$ progressively decreases as the ratio $E_J/E_C$ is reduced (see Fig.~\ref{fig:jump}), making the estimation of both the discontinuity and the critical point increasingly difficult. This suppression arises because, for small $E_J/E_C$, the variance at the transition becomes large, albeit finite.
Consequently, even though the symmetry between even and odd minima is broken and the system prefers the minimum at the origin, the wavefunction spreads significantly over adjacent minima; this broadening results in a marked reduction of $m^2$.

\section{Appendix C: sub-Ohmic and super-Ohmic dissipation}
\begin{figure}[h]
\centering
\includegraphics[width=\columnwidth]{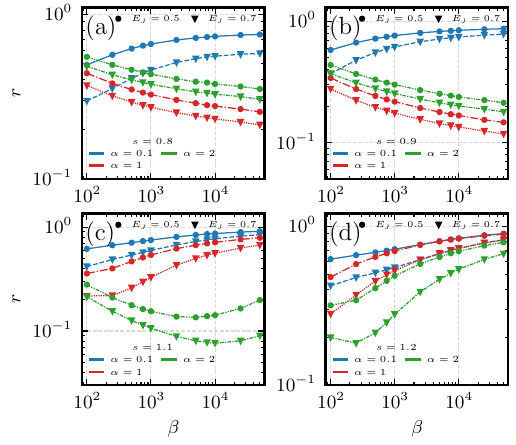}
\caption{Ratio $r$ between $\sigma^2$ and $\sigma_0^2$ at $E_J/E_C = 0.5$ (squares) and $E_J/E_C = 0.7$ (triangles), as functions of $\beta$ (in units of $1/E_C$). Specifically, data are shown for $\alpha = 0.1$ (blue), $\alpha = 1$ (red), $\alpha = 2$ (green), at sub-Ohmic $s = 0.8$ (a), and $s = 0.9$ (b), and super-Ohmic $s = 1.1$ (c), and $s = 1.2$ (d) dissipation. It is worth emphasizing that $r$ never approaches $0$ in the super-Ohmic regime, even for large $\alpha$, therefore the effect of the periodic potential does not change the low temperature behavior of the variance.}
\label{fig:svar}
\end{figure}

In this section, we show the absence of a QPT for any $\alpha>0$ for both the sub-Ohmic ($s<1$) and the super-Ohmic ($s>1$) regime, considering the spectral function~\eqref{eqn:spectral_form}. Indeed, the phase particle is expected to be always localized in the sub-Ohmic regime and always delocalized in the super-Ohmic regime, for any $\alpha>0$ and $E_J/E_C \geq 0$. Our Monte Carlo simulations confirm this scenario in a wide range of values of $\alpha$, and further show that the system properties are completely independent of the ratio $E_J/E_C$. Indeed, by comparing our numerical data with the analytical results for $E_J/E_C=0$, we find that the ratio $r = \sigma^2/\sigma_0^2 $ between the variance $\sigma^2$ computed at finite $E_J/E_C$ and the corresponding exact value $\sigma_0^2$ at $E_J/E_C=0$ converges to a finite but non zero value in the thermodynamic limit, for any $\alpha>0$ (see Fig.~\ref{fig:svar}). It proves that the introduction of a cosine potential does not affect the large-$\tau$ localization–delocalization behavior of the system. Specifically, in the range $1<s<2$, the $\beta^{s-1}$ behavior is still valid for non vanishing $E_J/E_C$. Although this analysis is performed in imaginary time, we expect the same qualitative features to hold in real time through analytic continuation.

\clearpage

\onecolumngrid
\begin{center}
\textbf{\large Supplemental Material for\\[0.5em] ``Quantum Brownian Motion: proving that the Schmid transition belongs to the Berezinskii–Kosterlitz–Thouless universality class''}\\[1em]
\end{center}

\setcounter{equation}{0}
\renewcommand{\theequation}{S\arabic{equation}}
\setcounter{figure}{0}
\renewcommand{\thefigure}{S\arabic{figure}}
\setcounter{section}{0}
\renewcommand{\thesection}{S\Alph{section}}

\section{The worldline Monte Carlo (WLMC) method}
In this section we explain the proposed WLMC scheme in detail. By using the path integral representation, the partition function associated with the model at $T=1/(k_B\beta)$ temperature is given by a sum over all periodic paths in imaginary time (cfr. Eq.~(2) in the main text):
\begin{equation}\label{eqn:partiton}
    Z = \int\limits_{\varphi(0) = \varphi(\beta\hbar)} {\cal D} [\varphi(\tau)] \,e^{-\mathcal{S}[\varphi(\tau)]/\hbar}.
\end{equation}
Subsequently, any observable $A$ can be represented in this framework as $A[\varphi(\tau)]$ allowing us to evaluate its thermal equilibrium average:
\begin{equation}\label{eqn:mean_value}
    \langle A \rangle = \frac{1}{Z}\int\limits_{\varphi(0) = \varphi(\beta\hbar)} {\cal D} [\varphi(\tau)] \, A[\varphi(\tau)]e^{-\mathcal{S}[\varphi(\tau)]/\hbar}.
\end{equation}
The WLMC method is a Markov chain Monte Carlo based technique and it involves the use of the trotterization of the action:
\begin{equation}\label{eqn:disc_action}\begin{split}
    \mathcal{S^{\text{dis}}}[\{\varphi_m\}] = -E_J \Delta \tau \sum_{i=0}^{N-1}\cos\varphi_i+
    \sum_{i<j} K^{\text{dis}}_{ij}(\varphi_i-\varphi_j)^2+\frac{\hbar^2}{4E_C \Delta \tau}\sum_{i=0}^{N-1}\bigl(\varphi_i - \varphi_{i+1}\bigr)^2,
    \end{split}
\end{equation}
being $N = \frac{\beta\hbar}{\Delta \tau}$ and $K^{\text{dis}}_{ij} = 2\Delta \tau^2 K\bigl(\Delta\tau (i-j)\bigr)$. The discretization process consists of replacing the differential $d\tau$ with a finite difference $\Delta \tau$ and substituting continuous paths $\varphi(\tau)$ with discretized paths $\{\varphi_m\}$. In order to guarantee the periodicity of the paths, we define $\varphi_{N} = \varphi_0$. The integer $i \in [0, N-1]$ denotes an imaginary time step $0\le\tau_i = i\Delta \tau <\beta\hbar$. In this context, the partition function becomes:
\begin{equation}\label{eqn:disc_part}\begin{split}
    Z^{\text{dis}} = \sum_{\{\varphi_m\}}e^{-\mathcal{S^{\text{dis}}}[\{\varphi_m\}]/\hbar} 
    \end{split}.
\end{equation}
Equivalently, the functional integral \eqref{eqn:mean_value} can be rewritten as:

\begin{equation}\label{eqn:disc_mean}\begin{split}
    \langle A^{\text{dis}} \rangle =\frac{1}{Z^{\text{dis}}} \sum_{\{\varphi_m\}}A(\{\varphi_m\})e^{-\mathcal{S^{\text{dis}}}[\{\varphi_m\}]/\hbar} 
    \end{split}.
\end{equation}
It is worth emphasizing that using \eqref{eqn:disc_part} and \eqref{eqn:disc_mean} instead of \eqref{eqn:partiton} and \eqref{eqn:mean_value} introduces an error of order $\Delta \tau$. The problem is now equivalent to a one-dimensional classical system of phase variables distributed on a chain of length $\beta$ and step $\Delta \tau$, interacting with each other. WLMC technique generates a Markov chain, which, after thermalization, samples the system's configurations $\{\varphi_m\}$, i.e. the worldlines, with the correct statistical weight, $p(\{\varphi_m\}) \propto e^{-\mathcal{S}^{\text{dis}}[\{\varphi_m\}]/\hbar}$, simplifying the calculation of \eqref{eqn:disc_mean}. We adopted an efficient sampling of the paths, a variant of the cluster algorithm proposed by Werner and Troyer \cite{Troyer_MC,Troyer_PRL}, based on Wolff algorithm \cite{Wolff_PRL}. Starting from a worldline $\{\varphi^{\text{old}}_m\}$, we randomly select a root $j$ as the first site of the cluster, along with a symmetry axis $a \in [-L\pi, L\pi]$. The update move consists of a reflection move with respect to the axis $a$ of any site $l$ belonging to the cluster as:
\begin{equation}\label{eqn:update}\begin{split}
    \varphi_l^\text{new} = 2a -\varphi_l^{\text{old}} 
    \end{split}.
\end{equation}
We build up the cluster by connecting any site $u$ of the worldline to an already added site $v$ with Wolff like probability:
\begin{equation}\label{eqn:prob_wolff}\begin{split}
    P_W(u|v)  = \max\biggl\{0, 1 -e^{-[S_{R}(\varphi^{\text{old}}_u,\varphi_v^{\text{new}})-S_R(\varphi^{\text{new}}_u,\varphi_v^{\text{new}})]/\hbar} \biggr\},
    \end{split}
\end{equation}
where we have defined the retarded interaction as
\begin{equation}\label{eqn:ret_action}\begin{split}
    S_R(\varphi_u,\varphi_v) =  \biggl( K^{\text{dis}}_{uv} +  \frac{\hbar^2}{4E_C\Delta \tau} (\delta_{v,u+1} + \delta_{v,u-1}) \biggr)(\varphi_u-\varphi_v)^2. 
    \end{split}
\end{equation}
Then we accept the reflection move of the whole cluster with a metropolis like probability:
\begin{equation}\label{eqn:prob_metr}\begin{split}
    P_M\bigl(\{\varphi_m^{\text{new}}\} \to \{\varphi_m^{\text{old}}\}\bigl)  = \min\biggl\{1,  e^{-[S_{J}(\{\varphi_m^{\text{new}}\})-S_{J}(\{\varphi_m^{\text{old}}\})]/\hbar} \biggr\},
    \end{split}
\end{equation}
where $S_J\bigl(\{\varphi_m\}\bigr) = -E_J\Delta\tau\sum_{i=0}^{N-1}\cos\varphi_i$ is the static interaction along the worldline. The algorithm we exploited is made up of two micromoves. The first one requires the axis to be chosen among the symmetry points of the Josephson potential: $a = k\pi$, where $k$ is an integer in the range $[-L, L]$ as proposed in the original work by Werner and Troyer \cite{Troyer_MC, Troyer_PRL}. Such choice of the axis ensures the acceptance probability \eqref{eqn:prob_metr} to be $1$. A more efficient option is to choose the axis among the three multiples of $\pi$ closest to the root. The use of this micromove alone does not guarantee the ergodicity of the algorithm, therefore we perform a second update where the axis is chosen randomly in the neighborhood of the root. Such micromove has a nonzero probability of generating single-site clusters. Therefore, since the reflection axes are randomly chosen, it ensures ergodicity, as every system configuration can be reached within a finite number of moves. The parameter $L$ should be large enough to contain the entire worldline, and it can be determined during the thermalization process. It is worth to stress that even $L$ being finite it will be sufficiently large that the boundaries no longer affect the system: this procedure is then equivalent to consider the extended picture of the phase variable. An important aspect to consider is that we define a Monte Carlo step as a sequence of micromoves, both with a random axis and located at potential minima or maxima, that on average attempt to update $N$ sites, i.e. the entire worldline. Because of the nature of the long-range interaction, the time for a complete Monte Carlo step becomes $O(N^2)$, thus preventing an efficient exploration of large values of $\beta$. To address this issue, the algorithm can be further improved by incorporating the approach proposed by Luijten and Bl\"{o}te \cite{Luijten1, Luijten2}. First, we define a function $F_{\left|u-v\right|} $ such that
\begin{equation}
    \begin{split}
        F_{\left|u-v\right|} \ge S_R(\varphi_u^{\text{old}},\varphi_u^{\text{new}}) - S_R(\varphi_u^{\text{new}},\varphi_u^{\text{new}}),
    \end{split}
\end{equation}
which depends only on $u$ and $v$, independently of $\varphi_u$ and $\varphi_v$. The cluster growth process is divided into two stages. The first stage involves the creation of provisional bonds with probability given by 
\begin{equation}\label{eqn:luijten_provv}\begin{split}
    P_{1LB}(u|v)  = 1 -e^{-F_{\left|u-v\right|}/\hbar}. 
    \end{split}
\end{equation}
This procedure can be implemented as follows: starting from a site $u$ we define the probability of forming no bonds from site $u+1$ up to $u+k-1$ and creating a bond between $u$ and $k$ as 
\begin{equation}\label{eqn:luijten_provv_up_to_k}\begin{split}
    P(k) = e^{-(F_1+F_2+\dots+F_{k-1})/\hbar}\biggl(1-e^{-F_{k}/\hbar}\biggr).
    \end{split}
\end{equation}
The associated cumulative probability is given by:
\begin{equation}\label{eqn:luijten_cum}\begin{split}
    C(k) = \sum_{k'=1}^{k} P(k').
    \end{split}
\end{equation}
No bonds will be formed with probability $1-C(N)$, while the first bond between $u$ and $u+k_1$ will be created with probability $C(k_1)-C(k_1-1)$. In the latter case we define the probability of forming no bonds from $u+k_1+1$ up to $u+k-1$ while creating a bond between $u$ and $k$ as
\begin{equation*}\begin{split}
    P_2(k) = e^{-(F_{k_1+1}+F_{k_1+2}+\dots+F_{k-1})/\hbar}\biggl(1-e^{-F_{k}/\hbar}\biggr) = \frac{P(k)}{1-C(k_1)}.
    \end{split}
\end{equation*}
The corresponding cumulative probability is:
\begin{equation}\begin{split}
    C_2(k) = \sum_{k'=k_1+1}^{k} P(k') = \frac{C(k)-C(k_1)}{1-C(k_1)}.
    \end{split}
\end{equation}
No further bonds will be formed with probability $1-C_2(N)$, while the second bond between $u$ and $u+k_2$ will be created with probability $C_2(k_2)-C_2(k_2-1)$. This process continues, iteratively connecting sites to $u$, until no more bonds are created. The same scheme is then repeated for each site added to the provisional cluster. The primary advantage of this approach is that, by employing a bisection algorithm, the bond selection per site can be performed in $O(\log N)$ time instead of $O(N)$ since $F_{\left|u-v\right|}$ is independent of the specific configurations of $u$ and $v$. In the second stage, each provisional bond is confirmed with probability
\begin{equation*}
\begin{split}
    P_{2LB}(u|v)  =  \frac{\max \biggl\{0, 1-e^{-[S_{R}(\varphi^{\text{old}}_u,\varphi_v^{\text{new}})-S_R(\varphi^{\text{new}}_u,\varphi_v^{\text{new}})]/\hbar}\biggr\}}{1 -e^{-F_{\left|u-v\right|}/\hbar}}. 
    \end{split}
\end{equation*}

\section{Renormalization Group approach}
Here, we provide a Renormalization Group (RG) argument to validate our numerical calculations and to demonstrate the absence of a quantum phase transition (QPT) at any finite value of $\alpha$ in the sub-Ohmic and super-Ohmic regimes. Our approach generalizes the framework of Ref.~\cite{Fisher_Zwerger} to the $s \neq 1$ regime. The resulting flow equations clearly evidence the fragility of the QPT when departing from the strictly Ohmic case ($s = 1$). 

We consider the Euclidean action for the Caldeira-Leggett model. Adopting natural units ($\hbar = 1$, $k_B = 1$), the action reads
\begin{equation}\label{eq:start}
    {\cal S}[\varphi] = \frac{1}{4E_C} \int_0^{\beta} d\tau\, \biggl(\frac{d\varphi}{d\tau}\biggr)^2 + \frac{\alpha \omega_c^{1-s}}{4\pi^2}\int_0^{\beta} d\tau \int_0^{\beta} d\tau' \frac{\bigl(\varphi(\tau)-\varphi(\tau')\bigr)^2}{|\tau -\tau'|^{1+s}} - E_J \int_0^{\beta} d\tau \, \cos \varphi(\tau),
\end{equation}
where the dissipative kernel is approximated by its asymptotic form for large $\tau$. This is justified since we are solely interested in the critical properties of the system, which are exclusively governed by its long-time (infrared) behavior.

Taking the zero-temperature limit ($\beta \to \infty$), we define the integral $\int d\tau \equiv \lim_{\beta \to \infty} \int_0^{\beta} d\tau$ and the Fourier transform $\varphi(\omega) = \int d\tau \, e^{-i\omega\tau} \varphi(\tau)$. The action can then be decomposed into a free and an interacting part, ${\cal S}[\varphi] = {\cal S}_0[\varphi] + {\cal S}_1[\varphi]$, with:
\begin{equation}
    {\cal S}_1[\varphi] = - E_J \int d\tau \, \cos \varphi(\tau),
\end{equation}
and
\begin{equation}
    {\cal S}_0[\varphi] = \frac{1}{2}\int_{-\omega_c}^{\omega_c} \frac{d\omega}{2\pi} \, \biggl(\frac{1}{2E_C}\omega^2 + 2F(\omega)\biggr)|\varphi(\omega)|^2 = \frac{1}{2}\int_{-\omega_c}^{\omega_c} \frac{d\omega}{2\pi} \, \biggl(\frac{1}{2E_C}\omega^2 + \frac{\alpha \omega_c^{1-s}}{\pi^2} A_s|\omega|^s\biggr)|\varphi(\omega)|^2.
\end{equation}
It is worth emphasizing that the high-frequency (ultraviolet) cutoff introduced to regularize the theory, is the natural cutoff defined in the main text at Eq.~(6).
Evaluating the Fourier transform yields the kernel $F(\omega) = 2\int d\tau \, \frac{\alpha \omega_c^{1-s} / (4\pi^2)}{|\tau-\tau'|^{1+s}} \bigl(1-\cos (\omega\tau)\bigr) = \frac{\alpha \omega_c^{1-s}}{2\pi^2} A_s|\omega|^s$. Here, $A_s$ is a positive constant in the physically relevant regime $0<s<2$, specifically defined as:
\begin{equation}
    A_s = 
    \begin{cases}
\frac{\pi}{2} & s=1 \\
-\Gamma(-s)\cos\biggl(\frac{\pi s}{2}\biggr) & 0<s<1 \,\, \text{or} \,\, 1<s<2.
\end{cases}
\end{equation}
Following Ref.~\cite{Fisher_Zwerger}, the kinetic (capacitive) contribution $\frac{1}{2E_C}\omega^2 |\varphi(\omega)|^2$ to ${\cal S}_0[\varphi]$ is neglected: in the regime $0<s<2$, the low-frequency behavior of the propagator is overwhelmingly dominated by the dissipative term, scaling as $\omega^{-s}$, rendering the kinetic term formally irrelevant in the infrared (IR) limit. Thus, we map the free action to:
\begin{equation}
    S_0[\varphi] \mapsto \frac{1}{2} \int_{-\omega_c}^{\omega_c} \frac{d\omega}{2\pi} \frac{\alpha \omega_c^{1-s}}{\pi^2} A_s |\omega|^s |\varphi(\omega)|^2.
\end{equation}
Following the standard RG procedure, we decompose the field $\varphi$ into slow ($<$) and fast ($>$) modes
\begin{equation}
    \varphi(\omega) = \varphi_<(\omega) + \varphi_>(\omega),
\end{equation}
where $\omega \in [-\omega_c/b, \omega_c/b]$ for the slow modes and $\omega \in [-\omega_c, -\omega_c/b]\, \cup \, [\omega_c/b, \omega_c]$ for the fast modes, with $b>1$.
Because ${\cal S}_0[\varphi]$ is purely quadratic, it perfectly decouples the fast and slow modes, resulting in ${\cal S}_0[\varphi] = {\cal S}_0[\varphi_<] + {\cal S}_0[\varphi_>]$. Consequently, the partition function can be expressed as follows:
\begin{equation}
    Z = \int {\cal D} \varphi \, e^{-{\cal S}[\varphi]} = \int {\cal D} \varphi_< \, e^{-{\cal S}_0[\varphi_<]} \int {\cal D} \varphi_> \, e^{-{\cal S}_0[\varphi_>]} e^{-{\cal S}_1[\varphi_< + \varphi_>]} \propto \int {\cal D} \varphi_< \, e^{-{\cal S}_0[\varphi_<]} \biggl \langle e^{-{\cal S}_1[\varphi_< + \varphi_>]} \biggr \rangle_>.  
\end{equation}
We define the effective slow-mode action ${\cal S}_{\mathrm{eff}}[\varphi_<]$ through the partial trace over the fast modes:
\begin{equation}
     e^{-{\cal S}_{\mathrm{eff}}[\varphi_<]} \equiv e^{-{\cal S}_0[\varphi_<]} \biggl \langle e^{-{\cal S}_1[\varphi_< + \varphi_>]} \biggr \rangle_>.  
\end{equation}
Expanding the interaction to first order in the cumulant expansion in $E_J$, the average yields:
\begin{equation}
      \biggl \langle e^{-{\cal S}_1[\varphi_< + \varphi_>]} \biggr \rangle_> = \biggl \langle e^{E_J\int d\tau \cos\bigl(\varphi_<(\tau) + \varphi_>(\tau)\bigr)}  \biggr \rangle_> = e^{E_J\int d\tau \bigl\langle\cos\bigl(\varphi_<(\tau) + \varphi_>(\tau)\bigr)\bigr \rangle_>} + O(E^2_J).  
\end{equation}
Using the trigonometric identity $\cos(A+B) = \cos A\cos B - \sin A\sin B$, we find:
\begin{equation}
    \bigl \langle \cos(\varphi_< + \varphi_>)  \bigr \rangle_> = \cos \varphi_< \langle \cos\varphi_> \rangle_> - \sin \varphi_< \langle \sin\varphi_>   \rangle_> =  \cos \varphi_< \langle \cos\varphi_> \rangle_> = \frac{1}{2}\cos \varphi_< \biggl\langle e^{i\varphi_>} + e^{-i\varphi_>} \biggr \rangle_>,
\end{equation}
where the term proportional to $\langle \sin\varphi_> \rangle_>$ vanishes identically because the unperturbed action ${\cal S}_0[\varphi_>]$ is even in $\varphi_>$. 
Applying Wick's theorem $\langle e^{i\varphi_>}\rangle_> = e^{-\langle\varphi^2_>\rangle_>/2}$, we obtain:
\begin{equation}
    \bigl \langle \cos(\varphi_< + \varphi_>)  \bigr \rangle_> = \cos \varphi_< e^{-\langle\varphi^2_>\rangle_>/2}.
\end{equation}
Since ${\cal S}_0[\varphi_>]$ is Gaussian, the variance $\langle \varphi^2_>\rangle_>$ can be computed exactly over the high-frequency shell:
\begin{equation}
    \langle \varphi^2_>\rangle_> = 2 \int_{\omega_c/b}^{\omega_c} \frac{d\omega}{2\pi} \frac{\pi^2}{\alpha \omega_c^{1-s} A_s \omega^s} = \frac{\pi}{\alpha A_s (1-s)} \bigl (1-b^{-(1-s)}\bigr) \equiv \Delta(b).
\end{equation}
Remarkably, the cutoff dependence $\omega_c^{1-s}$ cancels out exactly in the variance, confirming $\alpha$ as a consistent dimensionless running coupling. Substituting this result back gives $\bigl \langle \cos(\varphi_< + \varphi_>)  \bigr \rangle_> = e^{-\Delta(b)/2} \cos \varphi_<$, leading to the effective action:
\begin{equation}
    {\cal S}_{\mathrm{eff}}[\varphi_<] = \frac{1}{2} \int_{-\omega_c/b}^{\omega_c/b} \frac{d\omega}{2\pi} \frac{\alpha \omega_c^{1-s}}{\pi^2} A_s |\omega|^s |\varphi_<(\omega)|^2 - E_J e^{-\Delta(b)/2} \int d\tau \cos \varphi_<(\tau).
\end{equation}
To complete the RG step, we perform a scale transformation $\tau = b{\tilde \tau}$, which implies $\omega = b^{-1}{\tilde \omega}$. Because $\varphi$ is a dimensionless phase, it transforms as $\varphi_<(\tau) = {\tilde \varphi}({\tilde \tau})$ in the time domain, while its Fourier components scale as $\varphi_<(\omega) = b{\tilde \varphi}({\tilde \omega})$. Therefore, ${\cal S}_{\mathrm{eff}}[\varphi_<]$ is mapped onto a new action ${\cal S}'[{\tilde \varphi}]$:
\begin{equation}\label{eq:fin}
    {\cal S}'[{\tilde \varphi}] = \frac{1}{2} \int_{-\omega_c}^{\omega_c} \frac{d{\tilde \omega}}{2\pi}  \frac{\alpha \omega_c^{1-s}}{\pi^2} b^{1-s} A_s |\tilde \varphi(\tilde \omega)|^2 - E_J b e^{-\Delta/2} \int d\tau \cos \tilde \varphi (\tilde \tau).
\end{equation}
The action \eqref{eq:fin} has the same form of action \eqref{eq:start}, providing the renormalized coupling constants at scale $b$:
\begin{equation}\begin{split}
    \alpha (b) &= \alpha b^{1-s}\\
    E_J(b) &= E_J b e^{-\Delta(b)/2}.
\end{split}
\end{equation}
Taking into account an infinitesimal scale transformation $b = e^{dl} \approx 1+dl$, we expand the relevant quantities to linear order in $dl$: $b^{1-s} \approx 1 + (1-s)dl$, $\Delta(dl) \approx \frac{\pi}{\alpha A_s} dl$, and $b e^{-\Delta(dl)/2} \approx 1 + dl - \frac{\pi}{2 \alpha A_s} dl$. Introducing $\delta s = s-1$, we obtain continuous, cutoff-independent RG flow equations valid for $-1 < \delta s < 1$:
\begin{equation}
\begin{split}
    \frac{d \alpha}{dl} &= -\delta s \,\alpha,\\
    \frac{dE_J}{dl} &= \biggl(1-\frac{\bar{\alpha}_s}{ \alpha}\biggr)E_J,
\end{split}
\end{equation}
where $\bar{\alpha}_s = \frac{\pi}{2 A_s}$. Note that $\bar{\alpha}_s < 1$ for $s < 1$, $\bar{\alpha}_s > 1$ for $s > 1$, and $\lim_{s \to 1} \bar{\alpha}_s = 1$. 

In the non-Ohmic regime ($\delta s \neq 0$), the solution to these flow equations is as follows:
\begin{equation}
\begin{split}
    \alpha(l) &= \alpha(0)e^{-\delta s \, l},\\
    E_J(l) &= E_J(0) \exp \biggl[ l - \frac{\bar{\alpha}_s}{\alpha(0) \delta s} (e^{\delta s \, l} - 1) \biggr].
\end{split}
\end{equation}
In the Ohmic regime ($\delta s = 0$), the solution becomes:
\begin{equation}
\begin{split}
    \alpha(l) &= \alpha(0),\\
    E_J(l) &= E_J(0) e^{\bigl(1-\frac{1}{\alpha}\bigr)l}.
\end{split}
\end{equation}
These equations dictate the macroscopic fate of the system, which is highly sensitive to the spectral exponent $s$:
\begin{itemize}
    \item \textbf{Super-Ohmic regime ($\delta s > 0$):} The effective dissipation $\alpha$ scales to zero in the infrared limit ($d\alpha/dl < 0$). Consequently, the Josephson energy is ultimately suppressed. The system invariably flows toward a delocalized phase, precluding any quantum phase transition (QPT).
    \item \textbf{Sub-Ohmic regime ($\delta s < 0$):} The dissipation coupling $\alpha$ grows indefinitely under RG flow ($d\alpha/dl > 0$). Eventually, the effective Josephson energy diverges, indicating that the potential barriers become impenetrable. The system is therefore inherently localized for any non-zero bare coupling $\alpha(0) > 0$.
    \item \textbf{Ohmic regime ($\delta s = 0$):} Dissipation is strictly marginal ($d\alpha/dl = 0$). The flow of the Josephson energy reveals a phase transition that separates a delocalized ($\alpha < 1$, with $E_J(l\to \infty) \to 0$) and a localized ($\alpha > 1$, with $E_J(l\to \infty) \to \infty$) phase. As rigorously demonstrated by Bulgadaev \cite{Bulgadaev} by extending this analysis to the second order, the critical point precisely at $\alpha_c = 1$ is robust and holds independently of the bare ratio $E_J/E_C$.
\end{itemize}

\begin{figure}[h]
\centering
\includegraphics[]{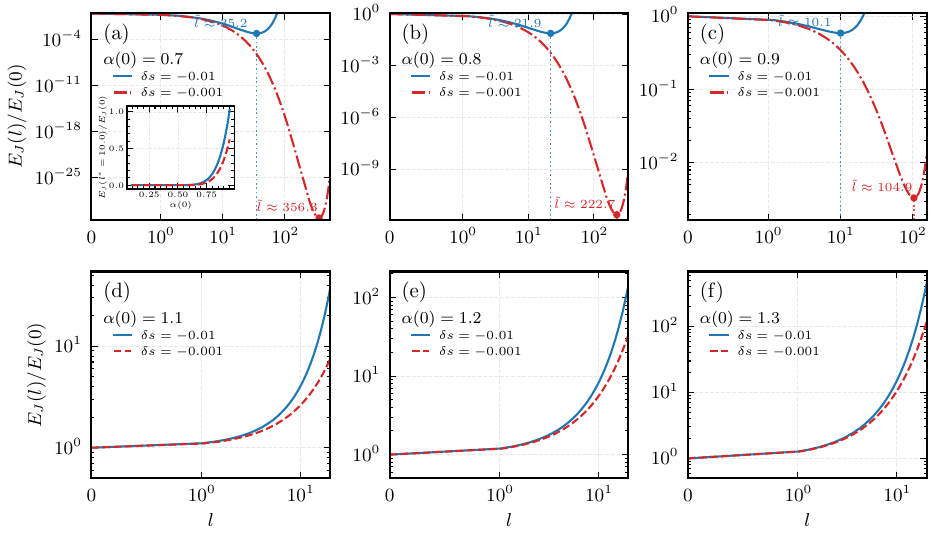}
\caption{Renormalized Josephson energy $E_J(l)/E_J(0)$ as a function of the RG time $l$ for different bare couplings $\alpha(0)$ in the sub-Ohmic regime, considering small deviations from the Ohmic case ($\delta s = -0.01$ and $\delta s = -0.001$). For $\alpha < \bar{\alpha}_s \approx 1$, the flow exhibits a minimum at $l = \tilde{l}$, which shifts to larger values as $\delta s \to 0$. This transient behavior may misleadingly suggest that the system is delocalizing for $\alpha < \bar{\alpha}_s$. Inset (a): $E_J(l^* = 10)/E_J(0)$ as a function of $\alpha(0)$. As shown, for a sufficiently short observation scale $l^*$, the renormalized coupling $E_J(l^*)$ is smaller than the bare value $E_J(0)$ across almost the entire $\alpha < \bar{\alpha}_s$ region.
}\label{Fig:dsneg}
\end{figure}

\begin{figure}[h]
\includegraphics[]{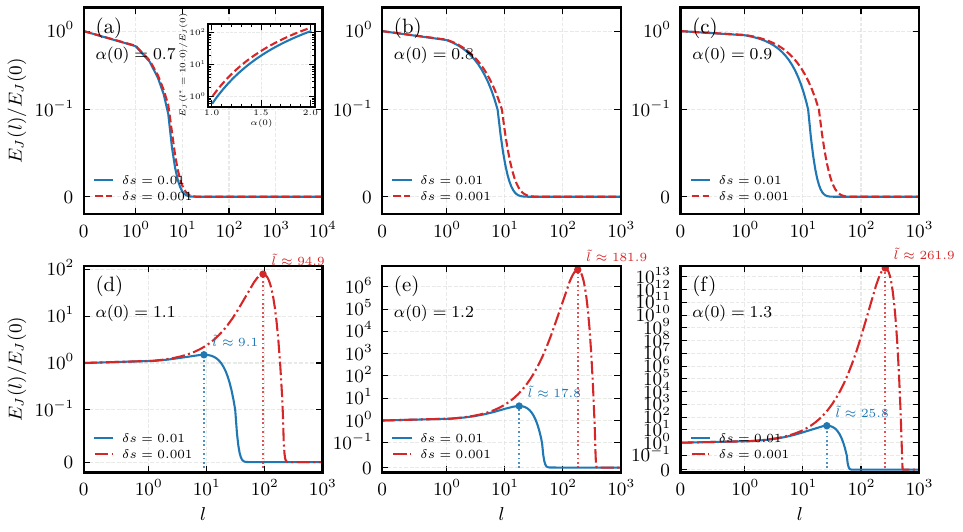}
\caption{Renormalized Josephson energy $E_J(l)/E_J(0)$ as a function of the RG time $l$ for different bare couplings $\alpha(0)$ in the super-Ohmic regime, considering small deviations from the Ohmic case ($\delta s = 0.01$ and $\delta s = 0.001$). For $\alpha > \bar{\alpha}_s \approx 1$, the flow exhibits a maximum at $l = \tilde{l}$, which shifts to larger values as $\delta s \to 0$. This transient behavior may misleadingly suggest that the system is localizing for $\alpha > \bar{\alpha}_s$. Inset (a): $E_J(l^* = 10)/E_J(0)$ as a function of $\alpha(0)$. As shown, for a sufficiently short observation scale $l^*$, the renormalized coupling $E_J(l^*)$ exceeds the bare value $E_J(0)$ across almost the entire $\alpha > \bar{\alpha}_s$ region.
}\label{Fig:dspos}
\end{figure}

Further analysis of the non-Ohmic regime reveals a stationary point $\tilde{l}$ in the evolution of $E_J(l)$:
\begin{equation}
    \tilde{l} = \ln \biggl(\frac{\alpha(0)}{\bar{\alpha}_s}\biggr)^{1/\delta s}.
\end{equation}
Since $l > 0$ by definition, the stationary point $\tilde{l}$ is physically accessible only when $\alpha(0) > \bar{\alpha}_s$ for $\delta s > 0$, and when $\alpha(0) < \bar{\alpha}_s$ for $\delta s < 0$. This extremum corresponds to a minimum (maximum) of $E_J(l)$ in the sub-Ohmic (super-Ohmic) regime (see Figs.~\ref{Fig:dsneg} and~\ref{Fig:dspos}). For a fixed $\alpha(0)$, the value of $\tilde{l}$ diverges as $\delta s \to 0$. 

Identifying the ultraviolet cutoff with the charging energy $E_C$ and the infrared observation scale with the physical temperature $T = \beta^{-1}$, the RG time is related to the energy scale via $l = \ln(\beta E_C)$. This mapping allows us to define a characteristic inverse temperature $E_C \tilde{\beta}(\alpha(0), s) = (\alpha(0)/\bar{\alpha}_s)^{1/\delta s}$. This characteristic scale highlights a finite-temperature window where the non-Ohmic environment is practically indistinguishable from an Ohmic one:
\begin{itemize}
    \item \textbf{Super-Ohmic regime ($\delta s > 0$):} For $\alpha(0) < \bar{\alpha}_s$, $E_J(l)$ decreases monotonically, clearly signaling a delocalized system. However, for $\alpha(0) > \bar{\alpha}_s$, $E_J(l)$ initially increases up to $\tilde{l}$. If the system is probed at a temperature such that $\beta < \tilde{\beta}(\alpha(0), s)$, this initial flow might erroneously suggest that the system is localizing, and then the presence of a QPT at $T=0$. In reality, $E_J(l)$ is a decreasing function for $l > \tilde{l}$, thus confirming the asymptotic delocalization for $\beta > \tilde{\beta}(\alpha(0), s)$.
    \item \textbf{Sub-Ohmic regime ($\delta s < 0$):} For $\alpha(0) > \bar{\alpha}_s$, $E_J(l)$ increases monotonically, signaling robust localization. In contrast, for $\alpha(0) < \bar{\alpha}_s$, $E_J(l)$ initially decreases up to $\tilde{l}$. Measurement of the system at temperatures $\beta < \tilde{\beta}(\alpha(0), s)$ might lead to the false conclusion that the system is delocalizing, and then the presence of a QPT at $T=0$. However, for $l > \tilde{l}$, $E_J(l)$ becomes an increasing function, ensuring localization in the true infrared limit $\beta > \tilde{\beta}(\alpha(0), s)$.
\end{itemize}

\end{document}